\documentclass[12pt]{article}

\usepackage{lmodern}
\usepackage[T1]{fontenc}
\usepackage[margin=1in]{geometry}
\usepackage[utf8]{inputenc}
\usepackage{graphicx}
\usepackage[authoryear,round]{natbib}
\setcitestyle{aysep={},citesep={;},yysep={,}}
\usepackage{amsmath}
\usepackage{amssymb}\usepackage{longtable}\newcommand{\E}{\mathbb{E}}\newcommand{\Prob}{\mathbb{P}}\DeclareUnicodeCharacter{0394}{\ensuremath{\Delta}}
\usepackage{amsthm}
\usepackage{bm}
\usepackage{newtxtext}
\usepackage[table,x11names]{xcolor}
\usepackage{booktabs}
\usepackage{tabularx}
\usepackage{array}
\usepackage{setspace}
\usepackage{enumitem}
\usepackage{microtype}
\usepackage{url}
\usepackage{placeins}
\usepackage{float}
\usepackage{caption}
\usepackage{bbm}
\usepackage{hyperref}
\usepackage{doi}

\newcommand{\AUC}{\mathrm{AUC}}
\newcommand{\ind}{\mathbbm{1}}
\newcolumntype{P}[1]{>{\raggedright\arraybackslash}p{#1}}

\newcolumntype{Y}{>{\centering\arraybackslash}X}
\hypersetup{
  colorlinks=true,
  citecolor=black,
  linkcolor=black,
  urlcolor=black,
  pdftitle={Punctuated Equilibrium in Technological Capabilities: Commercial Selection and the Predictability of Frontier Advances},
  pdfauthor={Kenzi M'Hamdi and Dawoon Jeong}
}
\begin{document}
\begingroup
\singlespacing
\begin{center}
\begin{spacing}{1}
{\fontsize{16}{20}\selectfont\bfseries
Punctuated Equilibrium in Technological Capabilities\par}
\vspace{0.3em}
{\fontsize{14}{18}\selectfont
Commercial Selection and the Predictability of Frontier Advances\par}
\vspace{1.2em}
{\large Kenzi M'Hamdi\(^{1}\)\quad Dawoon Jeong\(^{2,3*}\)\par}
\vspace{0.7em}
{\fontsize{10}{12}\selectfont\itshape
\(^{1}\) Brown University, Providence, RI, USA\par
\(^{2}\) Department of Sociology, University of Chicago, Chicago, IL, USA\par
\(^{3}\) Knowledge Lab, University of Chicago, Chicago, IL, USA\par
\vspace{0.25em}
\(^{*}\) Correspondence: Dawoon Jeong (\href{mailto:jdwoon0523@uchicago.edu}{\nolinkurl{jdwoon0523@uchicago.edu}})\par}
\end{spacing}
\end{center}
\vspace{0.5em}

\section*{Abstract}
Evolutionary accounts of technological change connect new technical opportunities with economic selection. Using approximately 1.74 million electronic-component product records, we examine the tempo of capability-frontier improvement alongside the diversity of commercially offered products. In a detailed analysis of 29 manually selected capabilities, 77.8\% of substantial-advance episodes consist of a single observed advance year. Across eligible three-year windows beginning with an advance, the initial observed gain accounts for an average of 79.3\% of total log-capability improvement. The median share of product records near the frontier is 1.46\%. Motivated by the interaction of technology push and demand pull, we extend the analysis to 204 capabilities and conduct forward out-of-sample comparisons between models using frontier history and models additionally incorporating product composition, manufacturer competition, and population dynamics. The additional indicators do not improve forecasts of subsequent frontier advances. The results document concentrated frontier gains alongside a broad distribution of commercially offered specifications. Stylized comparisons of incremental service value and switching costs identify candidate mechanisms linking application-specific valuation and production scale to the economic relevance of technical opportunities. They also specify the price, quantity, and application-level observations needed to evaluate these interpretations. We propose source-grounded, language-model-assisted extraction of demand-related information from dated external documents, with validation against independently observed market data. 

\vspace{1cm}
\noindent\textbf{Keywords:} punctuated equilibrium; technological capabilities; commercialized products; technology push and demand pull; evolutionary economics; forecasting.\par

\vspace{1cm}
\noindent\textbf{JEL classification:} O31; O33; L63.
\clearpage
\endgroup
\onehalfspacing

\section{Introduction}\label{sec:intro}
Evolutionary economics explains technological change through the generation of alternatives, the persistence and modification of routines, and selection under changing economic conditions \citep{nelsonwinter1982,dosi1994introduction,park2026quantifying}. This perspective directs attention to both the tempo of leading technical improvement and the diversity of products offered alongside it. A frontier advance can expand technical possibilities while a broad range of existing specifications continues to be offered commercially. The product-characteristics approach distinguishes these technical attributes from the services they provide \citep{saviotti1984theoretical,fontana2009product,windrum2009exploring}, while evolutionary accounts of variety treat the surrounding population as part of the process to be explained \citep{metcalfe1994competition,frenken1999variety}. Studying frontier trajectories and product diversity together therefore provides a basis for examining how new capabilities become economically relevant.

Punctuated equilibrium focuses attention on persistence interrupted by concentrated change \citep{gould1977punctuated,gersick1991revolutionary}. Research on technological discontinuities and step-like performance paths motivates measuring how much improvement occurs in brief advances and how those advances are distributed through time \citep{anderson1990technological,sood2005technological}. A long-run growth rate cannot distinguish repeated small gains from a similar total improvement concentrated in a short interval. Product-level evidence can make this distinction measurable across multiple technological functions. Two complementary quantities are particularly informative: the duration of episodes of substantial frontier advance, and the share of improvement supplied by the initial gain in windows beginning with an advance.

Technology push and demand pull motivate our comparison of frontier trajectories with the commercial product population. Technology push expands technical opportunities through accumulated knowledge and problem solving; demand pull shapes their direction and incentives through anticipated or realized needs and returns. Feedback from production and use connects these processes \citep{kline1986overview,dosi1982,vandenende2005technology}. Commercialized products---products offered for sale---are observable outcomes of that interaction. Their composition, manufacturers, and changing specifications may therefore provide signals of subsequent technical development. However, these features do not directly measure application-specific valuation, sales quantities, or the qualification and switching conditions under which buyers and producers respond to an improvement. Whether the observed population nevertheless improves forecasts beyond frontier history is an empirical question.

We investigate these questions using 1,739,040 historical electronic-component product records reconciled from DigiKey and Mouser distributor catalogs. The study makes three contributions. First, it brings large-scale product-level evidence to the measurement of punctuated technological change across diverse functions within electronics. In the detailed analysis of 29 manually selected anchor capabilities, 77.8\% of substantial-advance episodes consist of a single observed advance year. Across eligible three-year windows beginning with an advance, the initial observed gain accounts for an average of 79.3\% of total log-capability improvement. These statistics distinguish episode duration from gain concentration. Alongside this temporal concentration, the median share of product records near the frontier is 1.46\%, situating frontier improvements within a heterogeneous population of commercial offerings.

Second, the study assesses whether this surrounding population supplies useful advance signals. Extending the analysis to 204 capabilities, we compare models using frontier history with models additionally incorporating product composition, manufacturer competition, and population dynamics under the same forward out-of-sample evaluation. The additional indicators do not improve predictive performance. This comparison gives the study of technological trajectories a prospective dimension: it evaluates the incremental predictive information conveyed by observed commercial product characteristics. 

Third, we use comparisons of service value and switching costs to examine why products with different specifications remain commercially available. We develop stylized comparisons of incremental service value and switching costs to identify how application-specific benefits, production scale, and qualification or redesign conditions could shape the response to technical opportunities. These comparisons identify candidate mechanisms and the price, quantity, and application-level observations through which they could be evaluated. The resulting measurement agenda includes source-grounded, language-model-assisted extraction of demand-related information from dated external documents, with validation against independently observed market data. Linking these observations to product trajectories would allow subsequent work to assess both the proposed selection mechanisms and their contribution to forecasting further development.

Section~\ref{sec:literature} develops the literature and empirical expectations; Section~\ref{sec:methods} defines the data, measurements, and models; Section~\ref{sec:results} presents the results; and Section~\ref{sec:discussion} connects them to economic selection, candidate mechanisms, and the associated measurement agenda.

\section{Literature review}\label{sec:literature}
\subsection{Punctuated technological change}
Punctuated equilibrium originally concerns the relation between morphological persistence and comparatively rapid speciation in evolutionary history \citep{gould1977punctuated}. Its relevance to technological change lies in the distinction between the tempo of observed change and the process producing it. Applying the concept to capabilities requires identifying the measured attribute, the performance benchmark, and the temporal resolution at which an advance occurs.

Organization research emphasizes the alternation between relatively persistent structures and concentrated transformation. \citet{gersick1991revolutionary} compares change theories across several domains, while \citet{romanelli1994organizational} examine organizational transformations among minicomputer producers. At the technology level, \citet{anderson1990technological} distinguish technological discontinuities, competition among designs, and incremental development following the emergence of a dominant design. These contributions connect periods of concentrated change to the retention and reconfiguration of established arrangements.

Performance trajectories provide a complementary level of observation. Using 14 technologies from four markets, \citet{sood2005technological} find trajectories resembling steps separated by intervals of limited improvement rather than a single S-curve. Their evidence motivates explicit measurement of persistence, advance size, and the concentration of gains. Diffusion can generate a different temporal pattern: in the model of \citet{loch1999punctuated}, incremental technical improvements interact with externalities and resistance to switching to produce abrupt changes in use. Distinguishing the timing of technical benchmarks from the timing of population change is therefore important. Here, \emph{punctuated equilibrium at the capability level} denotes comparatively persistent performance benchmarks interrupted by concentrated advances.

\subsection{Product characteristics and the commercial product population}
The characteristics approach represents products through the technical features they embody and the services they provide \citep{saviotti1984theoretical,saviotti1996variety}. This allows technological outputs to be measured through their embodied capabilities. \citet{fontana2009product} positions detailed product databases within this research program, emphasizing data construction, variety, and the mapping from technical to service characteristics.

That mapping depends on combinations of attributes and the applications in which products are used. \citet{windrum2009exploring} show how groups of technical attributes support multiple service characteristics in mobile phones. The relevance of higher bandwidth, larger memory, or lower leakage current similarly depends on other attributes, operating conditions, interfaces, and intended use. An attribute-specific capability measure allows changes in these technical dimensions to be followed separately.

Population-level variety is equally important. \citet{frenken1999variety} use characteristics data on aircraft, helicopters, motorcycles, and microcomputers to examine variety and differentiation within product populations. In an evolutionary selection framework, diversity supplies the alternatives on which competition operates, and their changing distribution is part of technological change itself \citep{metcalfe1994competition}. A frontier benchmark and the population beneath it thus capture complementary aspects of the same evolving set of commercialized products.

More recent product-level studies reconstruct evolutionary relations from detailed characteristics. \citet{lee2024mapping} construct mobile-product phylogenies, \citet{jeong2024where} follow innovativeness along product lineages, and \citet{jeong2025mapping} infer connections between functional modules of successive products. These studies describe ancestry, recombination, and influence. Our analysis addresses the complementary relationship between the tempo of a numerical capability frontier and the position and movement of its commercial product population.

\subsection{Technology push, demand pull, and evolutionary selection}
Evolutionary economics separates the generation of alternatives through search from their selection and retention. Organizational routines support the persistence and modification of established practices, while heterogeneous firms and products carry different solutions \citep{nelsonwinter1982,dosi1994introduction,park2026quantifying}. The micro--meso--macro perspective places rules and their carrier populations at an intermediate level of analysis \citep{dopfer2004micro}. This directs attention to the distribution and persistence of commercialized alternatives alongside improvements at the frontier.

Technology push and demand pull describe interacting influences within this process. The chain-linked account of \citet{kline1986overview} connects technical knowledge, design, production, and market feedback. \citet{dosi1982} distinguishes problem-solving trajectories within technological paradigms from changes in the paradigms themselves. \citet{winter1984schumpeterian} examines how the location and accumulation of innovative knowledge alter competition between entrants and incumbents. Together, these perspectives relate the opportunities available to search to the conditions under which particular directions become attractive.

Demand helps shape those directions. The historical computing analysis of \citet{vandenende2005technology} relates technology-push and demand-pull arguments to different phases of paradigm development. \citet{tripsas2008} emphasizes discontinuities in customer preferences, \citet{chae2025warning} examines commercial pressure on technological trajectories, and \citet{schot2007niches} distinguishes niches and their evolutionary mechanisms. The economic relevance of an additional capability can therefore change with the application and selection environment even when the measured specification is unchanged.

Organizational and supply-chain conditions mediate this interaction. A component improvement can require changes in interfaces and architectural knowledge \citep{henderson1990architectural}; appropriability and complementary assets shape who captures its benefits \citep{teece1986profiting}; and exploration and exploitation allocate effort between new possibilities and established practices \citep{march1991exploration}. Commercial product features embody outcomes of these processes. Their numerical values may nevertheless be incomplete proxies for the demand, integration costs, and expected returns that influence subsequent development. This motivates treating the information content of those features as an empirical question.

\subsection{Frontier records, forecasting, and the information relationship}
Recent work by \citet{yinwang2026} studies frontier records across scientific and technical tasks in attempt order, reporting waiting-time regularities, sublinear record accumulation, and temporal dependence. Our observations instead follow products by recorded release year and construct annual capability benchmarks. This calendar-indexed representation allows frontier events to be compared with contemporaneous and lagged changes in the commercial product population.

Predictability also depends on the outcome being forecast. \citet{he2026benchmark} use technological phylogeny and strategic product features to forecast benchmark-product emergence. The present target is an annual capability-level advance defined by a thresholded increase in a cumulative quantile. The comparison between frontier-history and population-enriched models asks how much incremental predictive information the selected product indicators provide for that target.

The literature therefore motivates two connected tasks: characterizing frontier punctuation together with the surrounding product population, and evaluating the prospective information conveyed by that population. This distinguishes the economic connection between technological development and commercialization from the ability of a particular set of product features to reveal that connection. Prices, quantities, and application conditions provide a further set of candidate measurements through which the information relationship can be investigated.

\subsection{Research questions and empirical approach}\label{sec:questions}
The analysis asks how commercialized products are positioned relative to their capability frontier and how their distributions and movements relate to technical advance. The punctuated-change literature motivates the expectation that substantial capability gains concentrate in short episodes. The interaction between technical development and commercial selection motivates examining whether product-population indicators improve prediction beyond frontier history. Heterogeneity in selection conditions further motivates asking whether predictive performance or the benefit of population enrichment varies across capability characteristics and product classifications.

Episode-duration and gain-concentration statistics characterize the timing of substantial advances. Paired forward out-of-sample comparisons evaluate the expectation that product-population indicators add predictive value to frontier history. Regrouping the same out-of-fold predictions across capability characteristics and product classifications examines variation in predictive performance and enrichment benefit. Together, these analyses connect descriptive population evidence with prospective tests of the information represented by product features.

\section{Data and empirical methods}\label{sec:methods}
\begin{figure}[H]
\centering
\includegraphics[width=\textwidth]{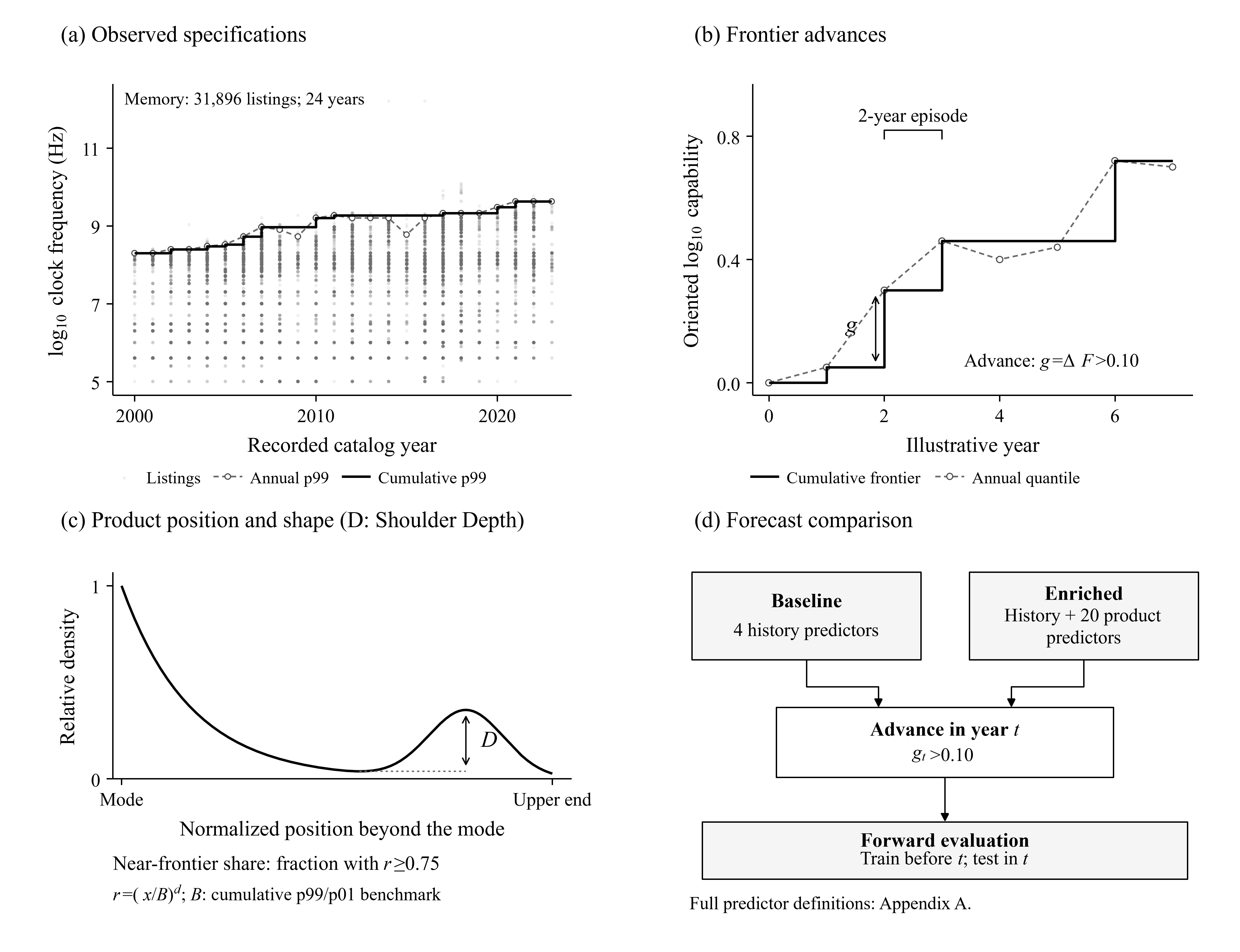}
\caption{\textbf{From commercial product trajectories to empirical variables.} a: memory-clock product records and annual and cumulative p99 benchmarks from the historical product database, restricted to usable years with at least 40 positive observations. Coincident points can overlap. b: schematic definitions of the annual quantile $f$, cumulative frontier $F$, gain $g$, event age $a$, inter-advance interval $T$, episode length $L$, and initial three-year gain share $c$. Episodes use calendar-consecutive advances; continuation shares use the following two calendar years. c: direction-adjusted product position, near-frontier share, the separate p95 lead ratio, and pooled-distribution shoulder depth $D$, defined as the height of a density rebound after a trough on the higher-capability side of the mode. d: the four frontier-history predictors and five groups comprising twenty additional product-population predictors. Panels b--d are schematic. Full predictor definitions are in Appendix~\ref{app:predictors}}
\label{fig:variables}
\end{figure}

Fig.~\ref{fig:variables} maps the empirical measurements, linking an observed memory-clock trajectory to schematic definitions of frontier change, product-population position, and forecasting inputs.

\subsection{Commercial product data, capabilities, and analysis samples}\label{sec:sample}
The historical product database reconciles DigiKey and Mouser distributor catalogs and contains 1,739,040 product records. A \emph{product record} is an observed listing for a commercialized component; population summaries count these records with equal weight. Cleaning harmonized field names, missing-value conventions, and numerical units, and coalesced complementary attributes. A \emph{capability} is a numerical product attribute within a specified subcategory, assigned a direction of improvement. The analysis combines a manually selected reference set of 29 capabilities used for detailed trajectory analysis, termed the \emph{anchor set}, with a broader screened sample. Selection for the broader sample required at least 1,000 records in a subcategory, 320 non-null observations for an attribute, and at least 0.15 log$_{10}$ change in a candidate direction, followed by semantic review. Of 907 candidate pairs, 661 pass coverage requirements, 722 pass the change requirement, and 556 pass both. The original selection contained 206 capabilities across 95 subcategories and 122 attribute columns. Measurement review withholds two capabilities pending repair: current rating in radio-frequency (RF) field-effect transistors (FETs), including metal--oxide--semiconductor FETs (MOSFETs), mixes leakage with current capacity, and operating supply current in low-dropout (LDO) voltage regulators includes a standby-current specification. The retained panel contains 204 capabilities across 95 subcategories and 121 attributes: 138 higher-is-better and 66 lower-is-better. The 29-capability anchor set is unaffected. A pin-specific current measurement in specialized power management remains flagged and is excluded in an additional sensitivity.

\subsection{Capability frontiers and population-relative measurements}\label{sec:frontier}
Let $x_{ijt}>0$ be the attribute value of product record $i$ for capability $j$ in usable year $t$. Let $d_j=+1$ for higher-is-better attributes and $d_j=-1$ otherwise. A usable capability-year has at least 40 positive, non-null observations, and a capability requires at least eight usable years. Define $Q_{jt}$ as the annual raw-value 99th percentile for $d_j=+1$ and the 1st percentile for $d_j=-1$. The oriented annual quantile and cumulative frontier are
\begin{equation}\label{eq:frontier}
 f_{jt}=d_j\log_{10}Q_{jt},\qquad
 F_{jt}=\max_{s\leq t}\{f_{js}\}.
\end{equation}

The raw-unit frontier is $B_{jt}=10^{d_jF_{jt}}$. A product record's direction-adjusted ratio to its own year's frontier is
\begin{equation}\label{eq:near}
 r_{ijt}=\left(\frac{x_{ijt}}{B_{jt}}\right)^{d_j}
 =10^{d_j\log_{10}x_{ijt}-F_{jt}},\qquad
 S_j(b)=\frac{\sum_{t\in\mathcal T_j}\sum_i\ind(r_{ijt}\geq b)}
 {\sum_{t\in\mathcal T_j}n_{jt}},
\end{equation}
where $\mathcal T_j$ contains usable years and $n_{jt}$ is their product-record count. We compute $S_j(0.75)$ within each anchor capability and summarize the 29 shares by their median. Alternative ratios are 0.90 and 0.50. The band is a direction-adjusted ratio. The lead-ratio illustrations and Appendix~\ref{app:leadpanel} use a cumulative p95 summary and a smoothed population median, adjusted for direction. Let $B^{95}_{jt}$ denote that raw-unit benchmark and $\widetilde M_{jt}$ the smoothed median. Their relative lead is
\begin{equation}\label{eq:lead}
 G^{95}_{jt}=\left(B^{95}_{jt}/\widetilde M_{jt}\right)^{d_j}.
\end{equation}
The lead ratios summarize within-capability start/end comparisons. Their p95 convention is distinct from the cumulative p99 benchmark in equation~\eqref{eq:near}. Appendix~\ref{app:provenance} documents implementation and provenance.

\subsection{Advance events, episodes, and concentration of gains}\label{sec:events}
Write $t_j^-$ for the preceding usable observation year. A frontier gain and thresholded advance are
\begin{equation}\label{eq:event}
 g_{jt}=F_{jt}-F_{j,t_j^-},\qquad
 A_{jt}(h)=\ind\{g_{jt}>h\}.
\end{equation}
The principal threshold is $h=0.10$ log$_{10}$ units, equivalent to an approximately 1.26-fold increase in oriented capability. Gains are measured between successive usable observation years, so the interval spanned by $g_{jt}$ depends on their calendar spacing.

An episode is a maximal run of calendar-consecutive observations with $A_{jt}=1$, broken by a missing year or a non-advance. With episode set $\mathcal E(h)$ and duration $L_e$ in observed consecutive years, the singleton fraction is
\begin{equation}\label{eq:single}
 P_1(h)=\frac{\sum_{e\in\mathcal E(h)}\ind(L_e=1)}{|\mathcal E(h)|}.
\end{equation}
We vary $h$ to describe sensitivity to the event definition. Episode size in the size--duration diagnostic is its maximum step, $\max_{t\in e}g_{jt}$. For an advance at $t$ with both following calendar years observed, the initial share of the ensuing three-year log-capability gain is
\begin{equation}\label{eq:gainshare}
 c_{jt}=\frac{g_{jt}}{g_{jt}+g_{j,t+1}+g_{j,t+2}}.
\end{equation}
We average this statistic over the eligible 73 windows and examine whether initial gain correlates with the subsequent share $1-c_{jt}$. Eligible windows begin with an advance and contain both following calendar years. A window is retained whenever these conditions hold, including when it overlaps another eligible window.

\subsection{Density rebounds in product distributions: shoulder analysis}\label{sec:shoulders}
In this study, a \emph{distributional shoulder} denotes a renewed rise in product density after a trough on the higher-capability side of the modal specification. The shoulder analysis measures the size of this rebound in each pooled capability distribution. Positive values are log$_{10}$-transformed and oriented, cropped at the 0.5th and 99.5th percentiles, and used in a 400-bin histogram smoothed with a Gaussian kernel and a Silverman-style bandwidth. Density estimation is capped at 20,000 observations. The interval between the mode and upper cropped endpoint is rescaled to $[0,1]$ and sampled at 60 positions; modal density is normalized to one.

Let $\widetilde p_j(u)$ be that normalized density and $u_{v,j}\leq0.88$ the trough selected by the estimator. \emph{Shoulder depth} is the subsequent density rebound,
\begin{equation}\label{eq:shoulder}
 D_j=\max_{u\geq u_{v,j}}\widetilde p_j(u)-\widetilde p_j(u_{v,j}).
\end{equation}
Shoulder depth is dimensionless. Its reference interval runs from the pooled mode to the upper cropped endpoint. Uncertainty is summarized using 1,000 part-level bootstrap samples. A separate comparison uses 1,000 samples from a fitted skew-normal unimodal distribution, repeating cropping and density estimation. We report three criteria: $D_j\geq0.04$; an observed bootstrap lower bound of at least 0.04; and an observed lower bound exceeding the null distribution's 97.5th percentile.

\subsection{Predictive specifications and forward evaluation}\label{sec:forecast}
The forecasting target is the indicator $A_{jt}$ for a candidate outcome year. The baseline input $\mathbf H_{j,<t}$ has four variables: recent cumulative-frontier change, recent change in the raw annual quantile, time since a previous advance, and lagged oriented frontier level. The implementation calls the last variable \emph{maturity}; throughout the analysis it denotes the lagged frontier level. Own-history changes use a three-observation lookback. Appendix~\ref{app:predictors} provides the predictor definitions.

The enriched specification adds $\mathbf Z_{j,<t}$: twenty variables in five groups, comprising four competitive-state variables, two sibling-capability indicators, two product-count/growth variables, nine population-motion variables, and three challenger variables. \emph{Sibling capabilities} are other capabilities in the same product subcategory. Manufacturer shares are calculated from product-record counts. A full variable dictionary links these groups to Fig.~\ref{fig:variables}.

With $\sigma(v)=1/(1+\mathrm{e}^{-v})$ and tildes indicating the reported median imputation and standardization, the two fitted prediction rules can be written as
\begin{align}
 \widehat p^{(0)}_{jt}
 &=\sigma\!\left(\widehat\alpha_0+
        \widehat{\bm\beta}_0^{\top}\widetilde{\mathbf H}_{j,<t}\right),\label{eq:baseline}\\
 \widehat p^{(1)}_{jt}
 &=\sigma\!\left(\widehat\alpha_1+
        \widehat{\bm\beta}_1^{\top}\widetilde{\mathbf H}_{j,<t}
       +\widehat{\bm\gamma}^{\top}\widetilde{\mathbf Z}_{j,<t}\right).
       \label{eq:enriched}
\end{align}
Both use $L_2$-penalized logistic regression with the reported software setting $C_{\rm logit}=1$. In generic notation, fitting balances binary log loss and a squared-coefficient penalty,
\begin{equation}\label{eq:penalty}
 \min_{\alpha,\bm\theta}\left\{
 \sum_{(j,t)\in\mathcal D_{\rm train}}
 \left[-A_{jt}\log p_{jt}-(1-A_{jt})\log(1-p_{jt})\right]
 +\lambda_{\rm pen}\|\bm\theta\|_2^2\right\}.
\end{equation}
Equation~\eqref{eq:penalty} expresses the penalized estimation principle; normalization and intercept treatment follow the software implementation.

For evaluation fold $Y$, the training set contains observations with outcome years strictly earlier than $Y$, and predictions are scored in $Y$. Training requires at least 30 rows and both outcome classes. The area under the receiver operating characteristic curve (AUC) additionally requires both classes in the test year. Nineteen scorable folds span 2005--2023. The earliest fits can contain only 55 training rows; an entirely missing predictor, including product-inflow acceleration in early folds, is dropped by the imputer. All predictor values precede the candidate outcome year.

Let $n_Y$ be the test sample size and define the paired annual difference
\begin{equation}\label{eq:aucdiff}
 \Delta_Y=\AUC_Y^{(1)}-\AUC_Y^{(0)},\quad
 \overline\Delta=\frac1{19}\sum_Y\Delta_Y,\quad
 \overline\Delta_w=\frac{\sum_Y n_Y\Delta_Y}{\sum_Y n_Y}.
\end{equation}
The confidence summaries resample annual folds 2,000 times to summarize uncertainty in the paired test-year differences. We also report retained-panel binned calibration comparisons, calibration slopes, and Brier scores (BS),
\begin{equation}\label{eq:brier}
 \operatorname{BS}=\frac1{N_{\rm test}}\sum_{(j,t)\in\mathcal D_{\rm test}}
 (A_{jt}-\widehat p_{jt})^2.
\end{equation}
Retained-panel gradient boosting provides a model-class sensitivity check, using scikit-learn's default GradientBoostingClassifier with seeds 0, 1 and 2 after training-fold median imputation, without a tuning search. Calibration pools the 19 scorable folds into ten equal-width probability bins. Within each nonempty bin, we compare the mean predicted probability with the observed advance frequency and report its observation count; empty bins are identified in the figure. Its slope is fitted by logistic regression with an intercept on predicted log odds, clipping probabilities to $[10^{-6},1-10^{-6}]$, with $C=10^6$.

As post hoc diagnostics of annual forecast variation, we tabulate training and test sample sizes, event counts, missingness in the twenty added predictors, and changes in the test sample's subcategory composition. We also recompute the mean paired AUC difference after omitting individual test years and the two years with the largest losses. These omissions summarize the existing annual scores without refitting the models.

\subsection{Timing, stratification, and capability-level coupling}\label{sec:diagnostics}
For advance dates $\tau_{j,k}$, complete inter-advance intervals are $T_{j,k}=\tau_{j,k+1}-\tau_{j,k}$. The retained 204-capability panel supplies 109 complete intervals, 21 capabilities with at least two intervals, and 43 adjacent-interval pairs. Diagnostics compare pooled and within-capability dispersion, an age--hazard slope, and adjacent-gap correlation with 2,000 within-capability event-position permutations (seed 0). The age--hazard statistic is a logistic coefficient on elapsed calendar years since the last observed event, using 1,186 rows with defined age. Within-capability dispersion is the unweighted mean of eligible capability-level coefficients of variation, each defined as the standard deviation divided by the mean. The full-window shuffle preserves event counts. An additional eight-year-block shuffle, with blocks $\lfloor(t-2000)/8\rfloor$, allows some slow rate variation and was introduced after the primary comparison. Boundary-censored waits are omitted, and null comparisons use the same observed support. Two-sided permutation probabilities include ties and double the smaller empirical tail, capped at one.

The stratified forecasting comparison uses both arms' regenerated forward out-of-fold predictions on the retained panel and regroups them by capability, category, subcategory, and product-record-volume strata. It includes 2,394 scored rows and 262 events across 2005--2024: the single-class 2024 test year contributes one prediction here and is omitted from the annual-AUC comparison. For the 160 capabilities with both outcome classes, shoulder depth is related to baseline AUC and enrichment benefit using Spearman correlation and 5,000 depth-label permutations (seed 0). Depth quintiles use capability-label order to break ties. Categorical cells require at least 40 rows, five events, and both classes. For each cut, 2,000 row-level label permutations (seed 7) preserve eligible cell sizes; maxima of baseline AUC and enrichment gain are evaluated separately. Predictions are regrouped after fitting the common models, and strata are constructed from full-sample product characteristics.

The coupling analysis measures co-movement between annual medians of the upper quarter and lower three quarters of product records. If $M^U_{jt}$ and $M^L_{jt}$ are those medians and $\mathcal D$ denotes the detrending operation, its correlation is
\begin{equation}\label{eq:coupling}
R_j=\operatorname{cor}\!\left(\mathcal D[\log M^U_{jt}],\,
                  \mathcal D[\log M^L_{jt}]\right).
\end{equation}
The analysis also summarizes changes in the lower group's median during years without a cumulative-frontier advance. These comparisons describe changes in the composition of observed product offerings.

\section{Results}\label{sec:results}
\subsection{The frontier and the product population remain distinct}\label{sec:respopulation}
\begin{figure}[H]
\centering\includegraphics[width=\textwidth]{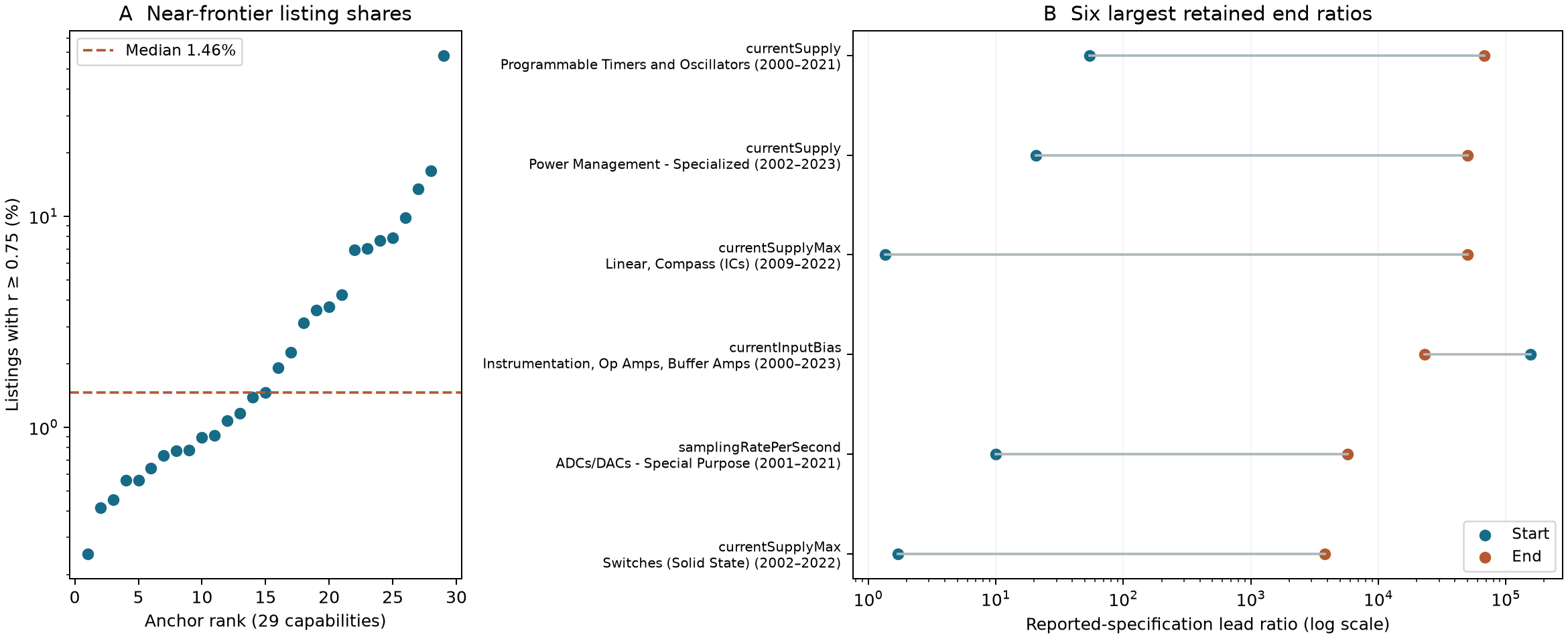}
\caption{\textbf{Frontier--population separation.} a: product-record shares near their own year's cumulative-p99 benchmark across 29 anchor capabilities; dashed line: median 1.46\%. b: the six largest retained end ratios of direction-adjusted cumulative p95 to the smoothed median, ranked by end ratio. Teal denotes each series' start and orange its end; years differ across series. Two confirmed incompatible current fields are withheld. The panels use different frontier summaries}
\label{fig:population}
\end{figure}

Fig.~\ref{fig:population}a shows the distribution of capability-specific near-frontier product-record shares in the 29-capability anchor set. The median is 1.46\% at the 0.75 ratio threshold. At ratios of 0.90 and 0.50 it is 1.29\% and 4.35\%, respectively. Most product records in a typical anchor capability therefore remain well below the cumulative benchmark, including under the wider of these bands.

Fig.~\ref{fig:population}b displays the six largest end ratios in the retained 204-capability panel, ranked by end ratio. Across all retained series, 179 ratios widen, 24 narrow, and one remains unchanged; the median within-series widening factor is 2.359. The full table appears in Appendix~\ref{app:leadpanel}. Smaller end ratios include optoisolator ratings near parity, showing heterogeneity in frontier--population separation. Appendix~\ref{app:trajectories} displays the original ratio-scale anchor trajectories, including concentrated steps, more gradual memory-clock improvements, and an essentially flat LDO output-current frontier.

The commercial product population thus occupies a broad range of positions below its leading performance benchmark. Frontier improvement and the distribution of offered specifications describe distinct dimensions of product evolution.

\subsection{Substantial advances concentrate in short episodes}\label{sec:respunctuation}
\begin{figure}[H]
\centering\includegraphics[width=\textwidth]{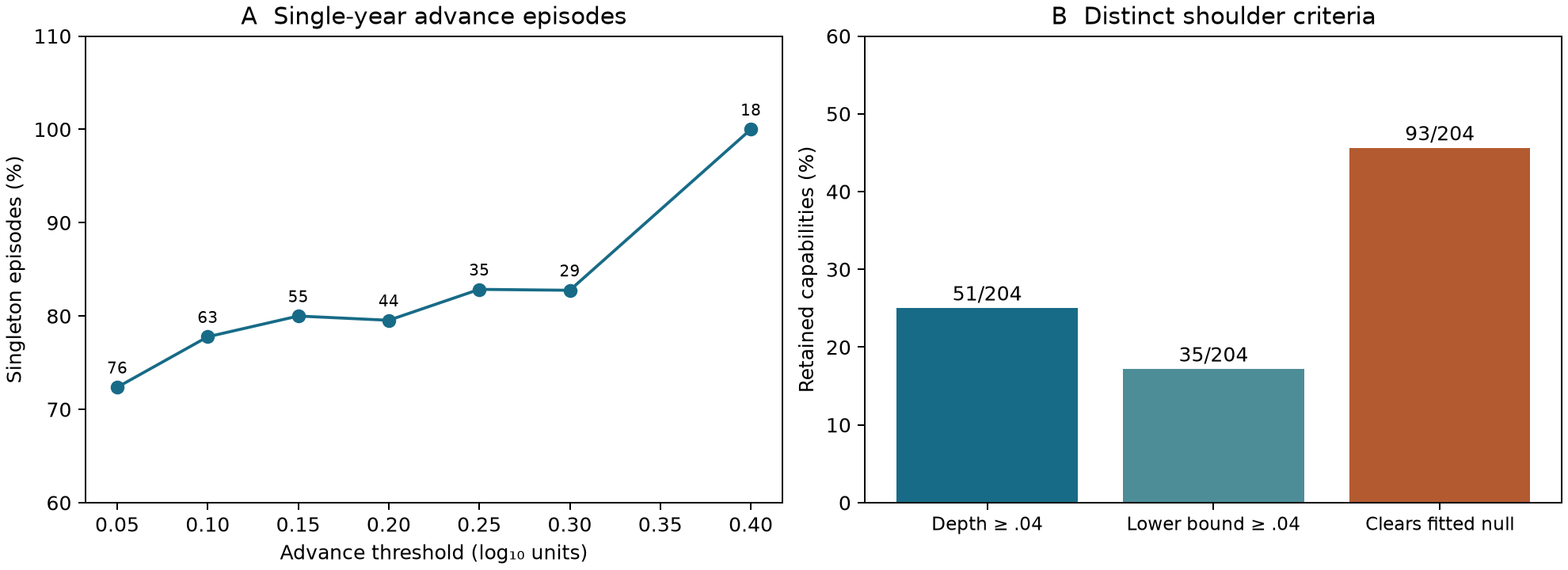}
\caption{\textbf{Punctuated advances and heterogeneous product distributions.} a: single-year episode share by advance threshold in the unchanged 29-capability anchor analysis; labels give episode counts. b: fractions of the retained 204 capabilities meeting the three shoulder criteria defined in Section~\ref{sec:shoulders}, obtained by restricting the existing capability-level estimates to the retained sample}
\label{fig:patterns}
\end{figure}

At $h=0.10$, 49 of 63 episodes consist of a single observed advance year: 77.8\%. Fig.~\ref{fig:patterns}a shows that the singleton share remains between 72.4\% and 82.8\% over thresholds 0.05--0.30. At 0.40, the 18 remaining episodes are all single-year. These findings quantify the contrast between relatively persistent frontier levels and temporally concentrated substantial changes in the selected anchors.

The continuation analysis provides a complementary measure of concentration. Across 73 advance-centered windows with both following calendar years observed, the initial observation supplies an average 79.3\% of the total three-year log-capability gain. Thus, within these windows, most improvement is concentrated at their beginning rather than accumulated evenly over the three observations. Table~\ref{tab:concentration} separates this result from the duration statistic and from the exploratory size--continuation comparisons.

\begin{table}[H]
\centering\small
\caption{Temporal concentration and continuation: distinct statistics and supports}\label{tab:concentration}
\begin{tabularx}{\textwidth}{P{0.50\textwidth}P{0.23\textwidth}Y}
\toprule
Quantity & Reported result & Support \\\midrule
Single-year share at $h=0.10$ & 49/63; 77.8\% & Anchor episodes \\
Singleton share, $h=0.05$--0.30 & 72.4--82.8\% & Threshold-specific episodes \\
Singleton share at $h=0.40$ & 18/18; 100\% & Remaining episodes \\
Mean initial share of three-year log gain & 79.3\% & 73 eligible windows \\
Peak episode gain vs. duration & $\rho_S=0.215$; $p=0.090$ & 63 episodes \\
Initial gain vs. following-two-year share & $\rho_S=-0.153$; $p=0.197$ & 73 windows \\
\bottomrule
\end{tabularx}
\end{table}

The size--duration and continuation comparisons yield the associations reported in Table~\ref{tab:concentration}. Taken together, the episode-duration and continuation results show that substantial capability gains concentrate in brief episodes.

\subsection{Product distributions exhibit heterogeneous density rebounds}\label{sec:resshoulders}
Of the 204 retained capabilities, 51 have estimated shoulder depth at least 0.04 (25.0\%), and 35 retain that depth at the bootstrap lower bound (17.2\%). A different comparison identifies 93 capabilities (45.6\%) whose observed lower bootstrap bounds exceed the fitted unimodal-null reference. Fig.~\ref{fig:patterns}b reports these three comparisons separately. These counts restrict the capability-specific estimates to the retained set. A full rerun of the original 206-capability bootstrap and null procedure reproduces the saved output exactly.

The results demonstrate variation in the shape of pooled product distributions. Some distributions exhibit a trough followed by renewed concentration toward the higher-capability end, while the extent and statistical prominence of this rebound differ across capabilities.

\subsection{Product-population indicators do not improve frontier forecasts}\label{sec:resforecast}
In the retained 204-capability panel, 2,449 capability-year rows contain 270 events; the 19 AUC-eligible test folds in 2005--2023 score 2,393 rows and 262 events. The enriched model's mean paired AUC difference is $-0.04167$, with a 2,000-resample fold-bootstrap interval of $[-0.07359,-0.01456]$. Weighting by test size gives $-0.03610$ and $[-0.06356,-0.01113]$. Fifteen of nineteen differences are negative. Rerun gradient-boosting comparisons at seeds 0, 1 and 2 give mean differences $-0.04043$, $-0.03875$ and $-0.04639$, respectively. Rebuilding the panel also recomputes sibling predictors without the excluded capabilities. Withholding the unresolved pin-specific current series as well gives a logistic mean difference of $-0.04405$ and interval $[-0.08020,-0.01542]$.

\begin{figure}[htbp]
\centering\includegraphics[width=\textwidth]{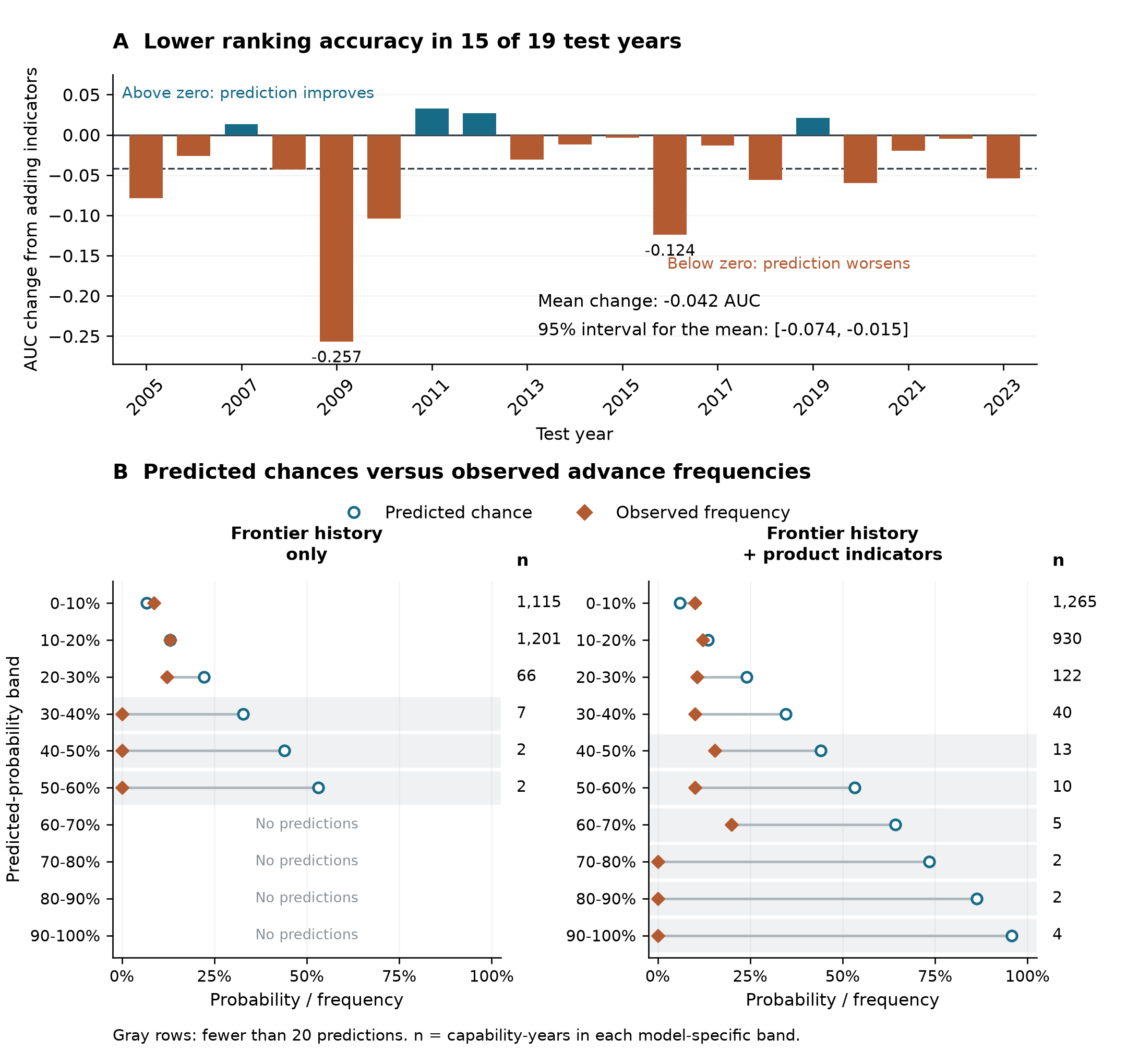}
\caption{\textbf{Adding product indicators lowers forecast accuracy in most years and yields overconfident higher-probability predictions.} a: annual AUC with twenty added product-population indicators minus AUC with four frontier-history predictors only; negative bars indicate poorer ranking. The dashed line is the mean; its 95\% interval resamples annual folds. b: within each model-specific probability band, open circles show mean predicted chance and filled diamonds show observed advance frequency. A diamond to the left of its circle indicates overprediction. Counts are capability-years; gray rows have fewer than 20. Both panels use the same 19 test years (2005--2023), comprising 2,393 predictions per model}
\label{fig:forecast}
\end{figure}

Annual differences are largest in 2009 and 2016. In 2009, the test sample contains 80 capability-years and only four advances; baseline and enriched AUCs are 0.539 and 0.283. All four advances rank below more non-advances under enrichment. The 2016 test sample contains 175 capability-years and eleven advances, with AUCs of 0.621 and 0.497. In that year, 31.7\% of entries in the twenty added predictors are missing, compared with 11.4\% in the training data, and the test sample spans 85 subcategories compared with 43 in 2015.

The negative average is not confined to these two years. Omitting 2009 yields a mean difference of $-0.0297$ across eighteen folds, with fold-bootstrap interval $[-0.0507,-0.0106]$. Omitting both 2009 and 2016 yields $-0.0242$ across seventeen folds, with interval $[-0.0426,-0.0067]$; thirteen of these seventeen differences remain negative. The mean remains negative after omitting any single test year.

Probability calibration also worsens. The baseline calibration slope is 0.436 with Brier score 0.0980, versus 0.057 and 0.1048 for the enriched model. Fig.~\ref{fig:forecast}b makes the probability comparison explicit: among the 122 enriched-model predictions in the 20--30\% band, the mean predicted chance is 24.0\% and the observed advance frequency is 10.7\%. In the 30--40\% band, these values are 34.6\% and 10.0\% across 40 predictions. Calibration pools predictions from the same 19 retained-panel folds.

Adding product-population indicators reduced discrimination and worsened calibration relative to the frontier-history baseline in forward testing.

\subsection{Timing and product-derived strata provide limited additional separation}\label{sec:restiming}
On the retained 204-capability panel, the within-capability inter-advance coefficient of variation is 0.793, with full-window permutation $p=0.006$ and eight-year-block $p=0.032$. The age--hazard slope is $-0.098$, with corresponding $p$-values 0.048 and 0.716. At the four-test threshold of 0.0125, one full-window comparison departs from the reference, but none of the block-controlled statistics does. Restricting to 2006 onward leaves 98 complete intervals and a within-capability coefficient of variation of 0.818 ($p=0.002$ under the full-window shuffle). The timing evidence therefore varies with the temporal reference model.

Among the 160 eligible retained capabilities, shoulder depth correlates 0.099 with baseline AUC (two-sided permutation $p=0.218$; directional $p=0.110$) and 0.015 with the enrichment benefit (two-sided $p=0.847$). No eligible category, subcategory or product-record-volume stratum clears its maximum-statistic comparison: baseline-AUC $p$-values are 0.984, 0.775 and 0.703, and enrichment-gain $p$-values are 0.804, 0.516 and 0.105, respectively. A supplementary correlation with enriched-model AUC is 0.144 (two-sided $p=0.072$; nominal directional $p=0.037$). The primary comparisons provide no support for systematic variation in baseline predictive performance or enrichment benefit across the tested capability characteristics and product strata.

\subsection{Population coupling differs across capabilities}\label{sec:rescoupling}
The case comparisons show that a persistent frontier can coexist with changes within the product population. For field-programmable gate array (FPGA) logic cells, detrended log medians of the upper quarter and lower three quarters of offerings correlate approximately 0.70 over 24 observations. Yet in nine years without cumulative-frontier advance, the lower-group median change averages $-0.195$ log$_{10}$ units. These years combine a stable frontier with a downward shift in the typical specification of the lower group.

Direct-current-to-direct-current (DC--DC) switching frequency also shows strong co-movement, approximately 0.76, but the average lower-group change in frontier-stall years is slightly positive. Across the original anchor set, 37\% of 494 stall observations have a falling typical value, while the median change is zero. The direction of population movement varies across capabilities. Optoisolator isolation ratings add a different example: values concentrate at a limited set of recurring ratings. Together, these cases distinguish movement at the benchmark, movement within the population, and concentration around established specifications.

\FloatBarrier
\section{Discussion}\label{sec:discussion}
\subsection{Product-level evidence on punctuated change and commercial heterogeneity}\label{sec:contributions}
Using approximately 1.74 million electronic-component product records, this study brings large-scale product-level evidence to evolutionary questions about the tempo of technological change and the coexistence of commercial alternatives. Its contribution links three observable dimensions: temporal concentration of capability-frontier advances, heterogeneity in the surrounding product population, and the prospective signals that this population supplies about subsequent technical advance.

First, in the detailed analysis of 29 selected anchor capabilities, 49 of 63 substantial-advance episodes (77.8\%) consist of a single observed advance year. Across 73 eligible three-year windows beginning with an advance, the initial observed gain accounts for an average of 79.3\% of total log-capability improvement. These measurements quantify the temporal concentration of frontier gains across diverse technological functions within electronics. They extend the characterization of step-like technological trajectories \citep{sood2005technological} by distinguishing episode duration from the share of improvement concentrated in the initial observed gain.

Second, the median near-frontier product-record share across the anchor capabilities is 1.46\%. The gap comparisons, heterogeneous density rebounds in the expanded sample, and capability-level coupling analyses show that frontier improvement occurs alongside a broad range of commercially offered specifications. Studying that range connects the product-characteristics approach to evolutionary accounts of variety and population change \citep{saviotti1984theoretical,metcalfe1994competition,frenken1999variety}. The frontier records leading performance in a particular attribute, while the population describes the commercial alternatives with which it coexists.

Third, the retained 204-capability panel supports a forward out-of-sample assessment of whether those alternatives convey useful signals of subsequent frontier advances. Adding twenty product-population indicators, including product composition, manufacturer competition, and population dynamics, to four frontier-history variables reduces mean paired fold AUC by 0.042, while probability calibration also deteriorates. The primary stratified comparisons identify no systematic differences across the tested capability characteristics and product groups. 

 Explaining these patterns requires measurements of application-specific valuation, production scale, and switching conditions.

\subsection{Technology push, demand pull, and the observability of selection}\label{sec:pushpull}
Evolutionary accounts of technology push and demand pull motivate the prospective comparison. Technical opportunities influence what can be developed, while expected demand and returns help determine the problems to which search is directed. Feedback from production and use can alter both the direction of technical development and the value attached to existing capabilities \citep{kline1986overview,dosi1982,vandenende2005technology}. If changes in the commercial product population reflect some of these influences before the frontier moves, population indicators may provide useful advance signals. The forecasting analysis evaluates this possibility for the available representation of commercial products.

Product features record outcomes of development and commercialization, but they provide an incomplete view of the conditions under which selection operates. A specification describes a capability embodied in an offered product. Its value for further development can depend on how many buyers require it, the additional service it provides, the volume over which a redesign cost can be recovered, and the qualification burden attached to replacement. Similar specification distributions can consequently be associated with different expected returns across applications and production scales. Changes in customer preferences can also redirect the importance of a capability \citep{tripsas2008}.

The negative forecasting result concerns these measured product indicators and the fitted models. Technology-push and demand-pull accounts allow feedback to operate through prices, quantities, use conditions, and expected returns that are absent from the present data. Moreover, the observed frontier is itself constructed from commercialized products and is already shaped by development and commercialization decisions. Both leading performance and the wider product population belong to this innovation process; their different observed movements do not establish a causal separation between technical development and economic selection.

This interpretation motivates an explicit account of the mechanisms connecting technical opportunities to economic incentives. Models of application-specific service value, production scale, and qualification or switching costs can specify when a feasible improvement is worth developing, incorporating, or adopting. Linking those models to observations of prices, quantities, and use would allow their implications for product persistence, uptake, and subsequent technical development to be evaluated. Such work would address the economic channels behind the observed trajectories while separately testing whether their measurement improves prediction.

\subsection{Adoption value, switching costs, and supply-chain value capture}\label{sec:threshold}
The combination of concentrated frontier gains and persistent product variety raises a mechanism-based question: under what conditions does an available technical improvement become valuable enough to develop, incorporate, or adopt? We compare the incremental service value of an improvement with its acquisition and switching costs. Attribute valuation distinguishes technical characteristics from buyer benefits \citep{rosen1974hedonic,lee2004estimating}, while switching-cost, replacement, and irreversible-investment research relate the value of changing an incumbent choice to the value of waiting \citep{klemperer1995switching,rust1987optimal,dixitpindyck1994}. 

For actor $a$, let $b_a(\mathbf z,r)$ denote annual monetary service benefit net of operating costs, and let $\mathcal C_a$ be a one-time incremental acquisition and switching cost, net of immediate disposal proceeds. The value of replacing an incumbent $\mathbf z^o$ now with $\mathbf z^n$, relative to retaining the incumbent over horizon $T$ with equal terminal values, is
\begin{equation}\label{eq:wtp}
 W_a(t)=\E_t\!\left[\int_0^T \mathrm{e}^{-\rho_a s}
 \{b_a(\mathbf z^n,r_{t+s})-b_a(\mathbf z^o,r_{t+s})\}\,\mathrm{d}s\right].
\end{equation}
The corresponding break-even comparison is $W_a(t)\geq\mathcal C_a$, relating the expected incremental service benefit of replacement to its acquisition and switching cost.

A restricted scalar illustration holds other attributes fixed and assumes initial incremental service flow $w\Delta q$, exponentially declining at rate $\delta$. Over an infinite horizon,
\begin{equation}\label{eq:threshold}
 W=\frac{w\Delta q}{\rho+\delta},\qquad
 \Delta q^*=\frac{(\rho+\delta)\mathcal C}{w}.
\end{equation}
Here $w$ is a local marginal annual service value and $\delta$ is the decay rate of incremental service. A higher service value can lower the required improvement conditional on costs, while larger switching costs raise it. With fixed redesign cost $\mathcal F$ spread over actual production volume $N$, a per-unit comparison has $\mathcal C=c+\mathcal F/N$. Demand pull enters through both the value of additional service and the scale over which fixed costs can be recovered. Technology push changes the feasible improvement $\Delta q$. The economic threshold $\Delta q^*$ thus links feasible capability gains to application-specific returns and costs; the empirical event threshold remains $h=0.10$.

These comparisons are made at different supply-chain stages. The component supplier evaluates development and qualification costs against retained incremental profit; the system producer evaluates integration and redesign; the user evaluates deployment and service benefits. Appropriability and complementary assets determine how value can be captured across these stages \citep{teece1986profiting}. An improvement beneficial to a user need not generate sufficient returns for its upstream developer, while a profitable supplier product need not justify every buyer's replacement. To explain continued commercial supply, one must observe producer incentives as well as users' reasons to retain an incumbent.

These comparisons give a candidate account of why substantial advances can coexist with the continued supply of products at established specifications. Existing capability may suffice for an application, while integration and qualification make replacement costly; in another application, additional capability may remove a valuable bottleneck. Historical discussions of long-lived electronic systems illustrate how redesign costs can exceed a part's procurement cost \citep{sandborn2008trailing}. Testing these channels requires matching capability gains to application-specific service benefits, actual production quantities, and the costs and returns faced at each supply-chain stage. Observing changes in uptake and continued supply alongside those conditions would help assess whether the proposed comparisons explain product persistence and the direction of further development. Appendix~\ref{app:interpretation} gives the associated finite-horizon, timing, and asset-value benchmarks.

\subsection{Managerial implications: irregular progress and response readiness}\label{sec:managerial}
A technology-monitoring system should distinguish the frontier, the position of typical commercialized products, and the time since the last substantial advance. These measures reveal different aspects of technological change. Their joint observation can be complemented by demand-side indicators: the applications in which improved capability creates value, the prices associated with that capability, and the expected volume over which development or integration costs can be recovered.

 The architectural-innovation literature explains why component changes can require interface and organizational adjustments \citep{henderson1990architectural}. Reusable qualification procedures, integration knowledge, and compatible designs may reduce the cost of responding to a useful advance. The economic assessment compares the cost of maintaining that readiness with the probability and value of a relevant future opportunity.

Periods of frontier persistence can also be used to assess where search and refinement are directed. Effort may support a different attribute, a new application, or improvements below the event threshold, consistent with the exploration--exploitation tension in organizational learning \citep{march1991exploration}. A product roadmap can distinguish these activities from preparations for a frontier advance and assess each against the applications and expected returns it serves.

\subsection{Implications for innovation-policy measurement and evaluation}\label{sec:policy}
For innovation policy, the frontier--population distinction highlights the need to measure capability creation, commercial availability, and realized use separately. The \emph{Oslo Manual} distinguishes innovation activities from implemented outcomes and includes activities that remain ongoing, are delayed, or are abandoned \citep{oecd2018oslo}. An evaluation that combines technical benchmarks with product variety, uptake, and service outcomes can better distinguish the stages through which innovation acquires economic significance.

Evaluation horizons matter when capability gains are concentrated in time. Short windows can miss infrequent advances, while longer windows admit more changes in technology, demand, and institutions. Pre-specified horizons and comparable unsupported units would allow policy evaluations to distinguish the timing of technical gains from their subsequent implementation and economic effects.

Research support, shared testing or qualification infrastructure, and demand-side instruments address different stages between technical possibility and use. The niche literature distinguishes selection environments and protected learning settings \citep{schot2007niches}. Their evaluation should identify the constraint addressed, the costs borne by different actors, and the effects on experimentation and implementation. Linking prices, quantities, and application evidence to technical capability records would make these distinctions more observable.

Such linkage would also make the distribution of economic benefits more visible. Technical improvements may affect suppliers, system producers, and users differently. A combined measurement system could examine who develops a capability, where it is taken up, and how its use changes service value or cost. Linkage quality, representativeness, commercial confidentiality, and the dates on which information becomes available are central design considerations.

\subsection{Observing economic selection: prices, quantities, and applications}\label{sec:context}
The stylized comparisons translate the interpretation into a measurement agenda. Incremental service value calls for observations of how capabilities are used and valued; production scale calls for quantities over which fixed costs can be recovered; and switching conditions call for qualification, compatibility, and redesign evidence. Product counts and manufacturer turnover describe commercial offerings and competitive presence. Matching these records to the proposed economic observations would help reveal the conditions that can redirect development before the recorded frontier moves \citep{dosi1982,vandenende2005technology,tripsas2008}.

Prices and quantities have complementary roles in this agenda. Matched transaction prices help characterize the valuation of product attributes under observed market conditions. Sales and production quantities describe the scale at which capabilities are exchanged and costs can be recovered. Application records indicate which additional services a capability can deliver, while qualification and integration records identify constraints on replacement. These observations require matching by product, application, supply-chain stage, and date, with attention to supply conditions and contracting. Table~\ref{tab:external} identifies candidate observations and the requirements for a subsequent study.

\begin{table}[htbp]
\centering\small
\caption{Candidate demand and selection-environment measurements for a subsequent study}\label{tab:external}
\begin{tabularx}{\textwidth}{P{0.24\textwidth}Y Y}
\toprule
Observation & Question it helps address & Measurement requirement \\\midrule
Matched transaction prices & How are attributes valued in traded products? & Match quality, transaction date, contract terms, and supply conditions. \\
Sales, orders and design wins & At what scale do products reach users and applications? & Distinguish unit sales, orders, and design selection; record their dates. \\
Installed use and utilization & Does extra capability deliver useful service? & Match workloads, service quality, users and dates. \\
Qualification and redesign records & What constrains switching? & Cost and delay are stage- and application-specific. \\
Standards and technical schedules & Which dated opportunities or constraints precede change? & Record publication dates separately from effective dates and later outcomes. \\
Service revenues and operating costs & Is replacement economically beneficial? & Measure incremental net service relative to the incumbent. \\
\bottomrule
\end{tabularx}
\end{table}

A subsequent study could evaluate these measurements against two connected questions: whether the proposed mechanisms account for product persistence and uptake, and whether the added observations improve forecasts of technical advance. For the predictive question, the same time-respecting frontier-history and product-population baselines could be compared with specifications augmented by economic measurements. The evaluation would distinguish exact-year advances, advances within a fixed horizon, cumulative capability gain, and uptake. Event discrimination, calibration, Brier score, and log loss would assess different aspects of performance. Economic identification would additionally require a design that distinguishes the proposed demand and switching channels from alternative supply-side explanations.

The information-theoretic identity in Appendix~\ref{app:information} clarifies the distinction between representing existing features and collecting additional evidence. Re-expressing product features can make them more useful to a fitted model. Independently measured context can also supply observations absent from those features. Comparing these routes under the same forward evaluation would help establish whether any predictive improvement comes from a better representation of existing records or from additional economic evidence.

\subsection{Source-grounded language-model extraction of selection conditions}\label{sec:llm}
Constructing these economic measurements at scale is a problem of extracting and linking evidence as well as obtaining data access. Dated procurement documents, manufacturer reports, application notes, and qualification records can contain statements about prices, orders, shipments, service requirements, and switching conditions. Large language models (LLMs) offer a possible means of recovering such statements and linking them to products, manufacturers, capabilities, and applications. Language-based extraction has already been used to construct chemical-function labels from patent text \citep{kosonocky2024mining}. We propose using LLMs to extract documented selection conditions that numerical product specifications omit.

The first task would be \emph{extraction}: identifying demand-related information explicitly present in dated external documents. Each extracted record would retain the supporting passage, source, publication and effective dates, unit, product match, and application context. Prices would be distinguished by transaction or quotation status, and quantities by orders, shipments, sales, or production. Independent human coding and independently observed market records would be used to assess whether the extracted statements are accurate, correctly matched, and economically comparable.

A separate task is \emph{estimation}, in which product and document representations are combined with observed market data to predict a missing economic variable. For prices, a hedonic model could relate observed transactions to numerical specifications, textual descriptions, and market context. \citet{bajari2025hedonic} show how language-model-based product representations and image features can be combined with observed prices to estimate hedonic prices. A component-level application could test whether textual application and compatibility information improves price estimation beyond numerical specifications alone. Estimates would retain their uncertainty and be explicitly distinguished from extracted observations.

Quantity estimation would require its own observed targets and validation. Verified order or shipment data could be combined with dated procurement information, manufacturer disclosures, and downstream application indicators where the evidence supports a product- or application-level match. Aggregated reports would support estimates at the corresponding level of aggregation; weakly documented cases could be represented by ranges or ordinal indicators. 

The proposed evaluation would first validate extraction and estimation against independent coding and held-out observed prices or quantities, assessing errors across product groups and time. It would then add only information available before each forecasting cutoff to the frontier-history and product-population specifications. Comparisons among numerical product features, textual representations of existing product records, and additional source-grounded economic observations would distinguish changes in representation from new evidence. Prospective evaluation would test whether these measurements improve frontier forecasts and whether their relation to uptake and persistence is consistent with the proposed economic mechanisms.

The exploratory pilots in Appendix~\ref{app:llmpilots} concern retrospective elicitation of frontier advances and external events. The source-grounded measurement and independent-validation design proposed here is a subsequent research agenda, and improved price, quantity, or frontier prediction remains to be demonstrated. Its purpose is to make application-specific valuation, production scale, and switching conditions observable enough to evaluate how economic selection interacts with punctuated technical development.

\subsection{Scope, limitations, and identification}\label{sec:limits}
\subsubsection{Population coverage and measurement}
The analysis concerns observed records of commercialized products, rather than a sales-weighted population or an installed base. The 29 anchor capabilities are manually selected, the original 206-capability screen is retrospective, and the retained forecasting panel contains 270 events, of which 262 enter the AUC-eligible evaluation. Near-frontier shares therefore describe the selected product records rather than aggregate electronics-market shares or adoption rates. Historical coverage, release-year provenance, missingness, variants, numerical units, and operating or test conditions can affect both frontier and population measures. Individual dates and extreme specifications have not all been independently validated. Two incompatible current fields are withheld, and the specialized power-management current series remains flagged for a pin-specific measurement, with its exclusion reported as a sensitivity. These checks do not validate every retained field. The frontier measures an attribute-specific benchmark among commercialized products, not the limit of experimental or scientifically feasible performance. Large specification ratios need not represent comparable engineering advantages when operating conditions or measurement definitions differ.

\subsubsection{Frontier position and economic interpretation}
A product's distance from the frontier in one attribute is not an overall ranking of quality or welfare. Other capabilities, reliability, compatibility, and application requirements may differ. The separation between frontier and population is not statistical independence, and a widening gap neither measures switching costs nor demonstrates that a buyer failed to adopt a superior product. The p95 start/end ratios and cumulative-p99 near-frontier shares use distinct benchmarks; their summaries should not be treated as estimates of a single common divergence process. Likewise, an endpoint gap does not establish temporal punctuation. Upper--lower population coupling measures co-movement between groups of offered products.

\subsubsection{Event construction and temporal concentration}
The cumulative-quantile convention creates persistent steps, and missing usable years can make an observed gain span several calendar years. A single observed advance year therefore need not mean that an improvement occurred within one calendar year. The analytical threshold $h=0.10$ is distinct from an economic adoption threshold. Episodes are defined using calendar-consecutive advances, and episode size is the maximum step rather than a sum that would mechanically tend to increase with duration. Higher thresholds retain fewer events. These statistics measure temporal concentration under the specified event definition. Establishing a punctuated-equilibrium mechanism requires comparisons with smooth-progress and rare-event reference models. The 79.3\% statistic is conditional on an initial advance and the next two calendar years being observed; its first step may inherit an earlier coverage gap. Overlapping windows and repeated capabilities also affect the continuation comparisons. The statistic does not measure the share of full-history improvement generated by the largest few records. Those claims would require additional whole-history concentration measures and suitable null comparisons.

\subsubsection{Distributional shape and dependence}
Shoulder depth is a dimensionless density-rebound statistic relative to the pooled distribution, whose upper endpoint is distinct from the annual frontier. A rebound can arise from mixtures of years, applications, product variants, measurement conventions, or quantized reporting. It is not an identified buyer threshold or contemporaneous economic niche. The depth cutoff, bootstrap-bound criterion, and unimodal-null comparison answer different statistical questions and do not provide interchangeable estimates of prevalence. Part-level resampling does not eliminate manufacturer or variant dependence, and the capability-wise comparisons are not a multiple-testing-adjusted estimate of economically generated shoulders. The population-coupling and stall comparisons likewise leave the demand and engineering conditions behind the observed differences unidentified.

\subsubsection{Forecasting and the interpretation of information}
The forecasting comparison evaluates the selected product indicators within the specified samples and models. Sparse early training sets, entirely missing predictors, regularization, functional form, and imperfect proxies can all affect the comparison. The annual diagnostics show how sparse advance counts and changes in predictor availability and sample composition accompany some large losses. With only four advances in 2009, each advance carries one quarter of that year's AUC through its ranking against non-advances. The greater missingness and broader subcategory coverage in 2016 are plausible contributors to instability under training-median imputation and a common fitted model, but their causal contributions are not identified.  The implementation's maturity variable is a lagged frontier level, not a normalized maturity index; sibling capabilities share a subcategory rather than a verified engineering dependency; and manufacturer shares weight product records rather than sales. Observation-based lookbacks can differ from calendar-year lookbacks. The predictive models omit capability and year fixed effects and structural adoption parameters. Shared categories, repeated capabilities, overlapping training windows, and common shocks limit the independence of annual folds and the interpretation of their bootstrap intervals. High-probability calibration bins are sparse. Consequently, the information gap denotes limited incremental prediction from the tested feature representation, not zero mutual information or causal separation between technical development and commercialization. Nor does the result establish that omitted demand caused the deterioration or that adding demand information will necessarily improve forecasts.

\subsubsection{Timing and stratification references}
The event-position permutations depend on the reference model: full-window exchangeability is stronger than the additional eight-year-block restriction, and boundary-censored waits are excluded. These diagnostics do not fit Poisson or Erlang arrival processes. The stratified comparisons regroup existing predictions rather than refitting models within independently measured economic regimes, and their full-sample strata are indirect proxies for selection conditions. Maximum-statistic comparisons control selection within each cut, not jointly across all cuts. The supplementary correlation with enriched-model AUC is distinct from the enrichment benefit and is not corrected across all depth comparisons. Its nominal directional significance does not change the conclusion of the primary stratified comparisons. Row-level permutations also leave dependence within categories and capabilities relevant. Differences in scored support follow the requirements of each test, including whether both outcome classes are needed.

\subsubsection{Economic identification and language-model estimates}
Prices, sales, buyer holdings, and stage-specific costs are absent from the present analysis. The adoption, supply-chain, and asset-value comparisons are proposed interpretations rather than fitted explanations. Service value is distinct from purchase price, incremental-service decay from accounting depreciation, and actual production volume from product-record counts. The break-even comparison does not establish an optimal replacement date, and the illustrative costs in the appendix do not calibrate the product population. Prices and quantities themselves reflect supply, demand, quality, and contracting; posted prices differ from transactions and willingness to pay. A predictive improvement from those variables would not by itself identify a causal demand mechanism. The managerial and policy implications likewise do not estimate returns to flexibility, intervention effects, or an optimal evaluation horizon.

LLM-based price or quantity estimates require observed validation targets, explicit uncertainty, and a distinction between extracted facts and predictions.  Transformations of existing product features do not supply independently observed demand information, and quantity inference requires separate evidence from price estimation. Historical knowledge in pretrained models creates a leakage risk that omitting a year from a prompt cannot remove. Prospective evaluation or defensible information cutoffs, source verification, and validation against independent observations are therefore essential. The retrospective pilots in Appendix~\ref{app:llmpilots}, including sensitivity to a packaging placebo, do not validate the proposed price or quantity measurements or demonstrate improved frontier forecasting. 

\section{Conclusion}\label{sec:conclusion}
This study examines the relationship between technological progress and commercialized products by following leading performance and the surrounding product population separately. In the 29-capability anchor analysis, 77.8\% of substantial-advance episodes occupy one observed year, and the initial observation accounts for 79.3\% of three-year log-capability improvement in the eligible advance-centered windows. The median near-frontier product-record share is 1.46\%. Expanded analyses reveal heterogeneous product distributions and capability-specific coupling. The tested product-population enrichment does not improve frontier-history predictions.

These findings identify a substantial information gap in the observed relationship between frontier advance and the features of commercialized products.  Technical specifications may leave important information about attribute valuation, quantities, application requirements, and switching conditions unobserved. Linking these features to prices, sales, use, and qualification evidence is therefore an important next step. LLM-assisted extraction and estimation may help construct those measurements when outputs are source-grounded and validated against observed market data. The resulting research agenda connects punctuated technical advance to the economic conditions under which capabilities are developed, selected, and used.

\FloatBarrier
\section*{Statements and Declarations}
\noindent\textbf{Funding.} NA\par
\noindent\textbf{Competing interests.} The authors declare that they have no competing interests.\par
\noindent\textbf{Author contributions.} 
\textbf{Conceptualization:} K.M., D.J.; 
\textbf{Methodology, Investigation, Formal analysis:} K.M.; 
\textbf{Data curation:} D.J.; 
\textbf{Writing --- original draft:} K.M., D.J.; 
\textbf{Writing --- review \& editing:} D.J., K.M.; 
\textbf{Supervision:} D.J.\par
\noindent\textbf{Data availability.} The data supporting this study are available from the corresponding author, Dawoon Jeong, upon reasonable request at \href{mailto:jdwoon0523@uchicago.edu}{\nolinkurl{jdwoon0523@uchicago.edu}}.\par
\noindent\textbf{Code availability.} The analysis code is available at the following Git repository: \url{https://github.com/kenmham/apto-semiconductors-replication}.\par

\clearpage
\appendix
\numberwithin{equation}{section}
\numberwithin{table}{section}
\section{Additional forecast-predictor definitions}\label{app:predictors}
The enriched specification adds twenty predictors to the four own-history variables. All are evaluated before the candidate outcome year. The four competitive-state variables are the oriented log gap from the annual best to the top-decile boundary, the gap from the upper-tail edge (p95, or p05 in the opposite direction) to the median, the effective number of manufacturers among top-decile product records, and the leader's consecutive tenure in usable annual observations. These state variables use the preceding usable observation, whereas the market-motion and challenger variables are keyed to the preceding calendar year. Effective manufacturer count is the inverse sum of squared product-record shares within that subset.

Two sibling variables record whether another capability in the same subcategory advanced in the preceding year and the count of sibling advances in the preceding three calendar years. Two product-count variables measure log product-record count at the preceding usable observation and its log change from the observation before that; these need not be adjacent calendar years. Nine market-motion variables describe newly appearing top-decile manufacturers, turnover in that manufacturer set, the number of distinct leaders, change in the edge-to-median gap, movement of the upper-tail edge, change in the fraction within a factor of 1.25 of the annual best, movement of the median and top-decile boundary, and acceleration in log product-record inflow. Turnover is one minus the intersection-to-union ratio of manufacturer sets in adjacent windows; inflow acceleration compares two consecutive three-year changes and therefore needs six years of history. Other endpoint changes use a three-calendar-year separation when both endpoints are available. This annual-best fraction differs from the cumulative-p99 shortfall measure reported in Section~\ref{sec:frontier}. The upper-tail-edge variable is called a runner-up slope in the code, but it is a population quantile, not the trajectory of one runner-up firm.

The final three variables track the current runner-up manufacturer: its gap from the annual leading manufacturer's best capability, its own three-year improvement, and how much its gap to the contemporaneous leader has closed over that interval. These distinguish a specific challenger from movement in the population quantiles. Sparse early coverage limits the effective predictor set in the earliest training folds.

\clearpage

\section{Start--end lead ratios across the retained capability panel}\label{app:leadpanel}
Table~\ref{tab:leadpanel} reports the retained capability panel. The original 206-capability panel is retained as an audit comparison. Following
measurement review, two capabilities are withheld pending repair: current rating
in RF FETs/MOSFETs and operating supply current in LDO Voltage Regulators.
Source checks identify leakage versus capacity and standby versus operating-current
mismatches, respectively. The retained descriptive panel contains 204 capabilities.
The specialized power-management supply-current series remains flagged because a
benchmark value is pin-specific; a sensitivity excludes it as well.

Ratios use cumulative annual p95 (p05 for lower-is-better quantities) and annual
medians smoothed in log10 units with Gaussian bandwidth 1.2 observation positions.
This is a retrospective, two-sided smoother. First and last usable years differ
across capabilities. Ratios describe reported numerical specifications, not a
common economic benefit. In particular, ratios of gain values in dB are not
linear power-gain ratios. Source-year provenance remains unresolved.

Of the 204 retained series, 179 widen, 24 narrow, and 1 remain unchanged. The median within-series widening factor is 2.359; median start and end ratios are 3.370 and 9.936. Using unsmoothed medians, 165 widen and the median factor is 2.075. These are conditional descriptions, not tests of temporal punctuation.

\small
\begin{longtable}{p{0.46\textwidth}rrrr}
\caption{Retained lead-ratio panel. Values are rounded for display}\label{tab:leadpanel}\\
\toprule Capability; direction & Years & Start & End & Factor\\\midrule\endfirsthead
\toprule Capability; direction & Years & Start & End & Factor\\\midrule\endhead
\bottomrule\endfoot
currentSupply\newline Programmable Timers and Oscillators; lo & 2000--21 & 54.3 & 6.78e+04 & 1.25e+03\\
currentSupply$^{a}$\newline Power Management - Specialized; lo & 2002--23 & 20.6 & 5.04e+04 & 2.44e+03\\
currentSupplyMax\newline Linear, Compass (ICs); lo & 2009--22 & 1.36 & 5.03e+04 & 3.7e+04\\
currentInputBias\newline Instrumentation, Op Amps, Buffer Amps; lo & 2000--23 & 1.57e+05 & 2.32e+04 & 0.148\\
samplingRatePerSecond\newline ADCs/DACs - Special Purpose; hi & 2001--21 & 10.1 & 5.72e+03 & 568\\
currentSupplyMax\newline Switches (Solid State); lo & 2002--22 & 1.72 & 3.81e+03 & 2.22e+03\\
currentSupply\newline Power Supply Controllers, Monitors; lo & 2002--21 & 16.6 & 3.23e+03 & 195\\
currentLeakageIsOffMax\newline Analog Switches, Multiplexers, Demultiplexers; lo & 2000--22 & 9.17 & 2.54e+03 & 276\\
3dbBandwidth\newline Instrumentation, Op Amps, Buffer Amps; hi & 2000--22 & 2.34 & 2.38e+03 & 1.01e+03\\
operatingSupplyCurrent\newline Switching Voltage Regulators; lo & 2011--21 & 6.82 & 2.21e+03 & 325\\
propagationDelayMax\newline Comparators; lo & 2000--18 & 113 & 1.71e+03 & 15.2\\
powerMax\newline IGBT Modules; hi & 2006--22 & 6.95 & 1.67e+03 & 240\\
currentSupply\newline Instrumentation, Op Amps, Buffer Amps; lo & 2000--22 & 59.5 & 1.64e+03 & 27.6\\
rotationalLifeCyclesMin\newline Encoders; hi & 2002--21 & 3.33 & 1.51e+03 & 452\\
currentSupply\newline Signal Buffers, Repeaters, Splitters; lo & 2002--21 & 178 & 860 & 4.82\\
currentCathode\newline Voltage Reference; hi & 2000--21 & 9.72 & 640 & 65.9\\
currentSupply\newline Voltage Reference; lo & 2000--22 & 10.5 & 591 & 56.4\\
samplingRatePerSecond\newline Analog to Digital Converters (ADC); hi & 2000--23 & 262 & 583 & 2.22\\
slewRate\newline Instrumentation, Op Amps, Buffer Amps; hi & 2000--23 & 109 & 556 & 5.11\\
gainBandwidthProduct\newline Instrumentation, Op Amps, Buffer Amps; hi & 2000--23 & 41.5 & 441 & 10.6\\
powerMax\newline Single Bipolar Transistors; hi & 2000--22 & 94.8 & 306 & 3.23\\
ifForwardCurrent\newline Rectifiers; hi & 2002--22 & 7.44 & 302 & 40.6\\
resetDelayTime\newline Supervisory Circuits; lo & 2012--22 & 1.34 & 301 & 225\\
powerMax\newline FET, MOSFET Arrays; hi & 2002--22 & 2.59 & 267 & 103\\
currentSensing\newline Current Sensors; hi & 2002--23 & 24.9 & 220 & 8.82\\
clockFrequency\newline Flip Flops; hi & 2001--17 & 18.1 & 196 & 10.8\\
currentOnStateItRmsMax\newline SCRs; hi & 2002--22 & 77.4 & 180 & 2.33\\
currentQuiescentMax\newline Comparators; lo & 2000--21 & 228 & 173 & 0.76\\
ippPeakPulseCurrent\newline Transient Voltage Suppression Diodes - TVS; hi & 2002--22 & 4.37 & 170 & 39\\
currentOnStateItAvMax\newline SCRs; hi & 2002--22 & 59.6 & 147 & 2.47\\
quiescentCurrent\newline Switching Voltage Regulators; lo & 2011--22 & 1.1 & 143 & 129\\
voltageInputOffset\newline Instrumentation, Op Amps, Buffer Amps; lo & 2000--23 & 17.2 & 127 & 7.38\\
maxSurgeCurrent\newline Rectifiers; hi & 2002--22 & 8.28 & 122 & 14.8\\
currentSupplyMax\newline Voltage Regulators - Linear, Low Drop Out (LDO) Regulators; lo & 2000--22 & 27.8 & 114 & 4.1\\
totalRamBits\newline FPGAs (Field Programmable Gate Array); hi & 2000--23 & 1.33 & 112 & 83.9\\
numberOfLabsClbs\newline FPGAs (Field Programmable Gate Array); hi & 2000--23 & 5.34 & 106 & 20\\
millicandelaRating\newline Circuit Board Indicators, Arrays, Light Bars, Bar Graphs; hi & 2001--19 & 15 & 98.5 & 6.56\\
currentQuiescentIq\newline Voltage Regulators - Linear, Low Drop Out (LDO) Regulators; lo & 2000--23 & 10.2 & 88.3 & 8.66\\
channelToChannelMatchingΔron\newline Analog Switches, Multiplexers, Demultiplexers; lo & 2000--22 & 11.4 & 87.4 & 7.67\\
operatingSupplyCurrent\newline Digital Isolators; lo & 2006--22 & 6.83 & 76.9 & 11.3\\
frequencyMax\newline Application Specific Clock/Timing; hi & 2002--22 & 3.41 & 63.9 & 18.7\\
currentSupply\newline RF Amplifiers; lo & 2002--23 & 17.7 & 60.7 & 3.42\\
numberOfLogicElementsCells\newline FPGAs (Field Programmable Gate Array); hi & 2000--23 & 4.74 & 57.2 & 12.1\\
quiescentCurrent\newline LDO Voltage Regulators; lo & 2006--22 & 4.78 & 56.8 & 11.9\\
reverseRecoveryTimeTrr\newline Diode Arrays; lo & 2000--23 & 6.34 & 54.6 & 8.62\\
frequencyMax\newline Clock Generators, PLLs, Frequency Synthesizers; hi & 2000--22 & 14 & 54.1 & 3.88\\
jitter\newline Clock Generators \& Support Products; lo & 2010--22 & 1.1 & 53.5 & 48.5\\
3dbBandwidth\newline Analog Switches, Multiplexers, Demultiplexers; hi & 2000--22 & 2.06 & 49.5 & 24\\
frequency\_max\newline Programmable Timers and Oscillators; hi & 2000--17 & 21.5 & 46.8 & 2.18\\
loadCurrentRating\newline Solid State Relays - PCB Mount; hi & 2002--21 & 22.7 & 45.7 & 2.01\\
currentOutputChannel\newline Logic Output Optoisolators; hi & 2002--21 & 1.02 & 35.3 & 34.7\\
currentOutputChannel\newline LED Drivers; hi & 2004--22 & 8.66 & 35 & 4.04\\
currentDcForwardIfMax\newline LED Emitters - Infrared, UV, Visible; hi & 2002--22 & 1.58 & 33.1 & 20.9\\
irReverseCurrent\newline Zener Diodes; lo & 2002--22 & 19.6 & 32.7 & 1.66\\
numberOfGates\newline FPGAs (Field Programmable Gate Array); hi & 2000--20 & 3.41 & 31.2 & 9.15\\
currentMax\newline RF Diodes; hi & 2002--22 & 10 & 29.9 & 2.99\\
onStateResistanceMax\newline Analog Switches, Multiplexers, Demultiplexers; lo & 2000--22 & 16.5 & 28.7 & 1.74\\
frequency\_max\newline LED Drivers; hi & 2004--22 & 29.2 & 28.4 & 0.974\\
millicandelaRating\newline LED Indication - Discrete; hi & 2000--22 & 34.9 & 28.3 & 0.812\\
frequencyMax\newline Clock Buffers, Drivers; hi & 2002--22 & 6.86 & 27.8 & 4.05\\
pulsesPerRevolution\newline Encoders - Industrial; hi & 2006--22 & 5.19 & 27.8 & 5.36\\
clockFrequency\newline Memory; hi & 2000--23 & 9.49 & 25.8 & 2.72\\
impedanceMaxZzt\newline Single Zener Diodes; lo & 2000--22 & 10.5 & 25.8 & 2.45\\
pppmPeakPulsePowerDissipation\newline Transient Voltage Suppression Diodes - TVS; hi & 2002--22 & 2.5 & 25 & 10\\
outputCurrent\newline Switching Voltage Regulators; hi & 2010--22 & 5.17 & 23.6 & 4.57\\
highSideVoltageMaxBootstrap\newline Gate Drivers; hi & 2000--22 & 2.08 & 23.2 & 11.2\\
writeCycleTimeWordPage\newline Memory; lo & 2000--22 & 155 & 22.5 & 0.145\\
temperatureCoefficientTyp\newline Digital Potentiometers; lo & 2000--21 & 15.5 & 22.2 & 1.44\\
currentAverageRectifiedIo\newline Single Diodes; hi & 2000--23 & 158 & 22.1 & 0.14\\
maximumRiseTime\newline Digital Isolators; lo & 2008--22 & 1.79 & 21.2 & 11.8\\
maximumFallTime\newline Digital Isolators; lo & 2008--22 & 1.54 & 21 & 13.6\\
operatingSupplyCurrent\newline Supervisory Circuits; lo & 2012--22 & 0.978 & 20.7 & 21.2\\
powerOutput\newline RF FETs, MOSFETs; hi & 2007--22 & 11.3 & 20.5 & 1.81\\
rdsOnDrainSourceResistance\newline MOSFETs; lo & 2007--22 & 18.2 & 20.4 & 1.12\\
currentAverageRectifiedIo\newline Bridge Rectifiers; hi & 2002--22 & 7.75 & 19.9 & 2.57\\
currentOutputMax\newline Power Distribution Switches, Load Drivers; hi & 2000--23 & 13 & 19.8 & 1.53\\
accessTime\newline Memory; lo & 2000--23 & 6.54 & 19.3 & 2.95\\
zzZenerImpedance\newline Zener Diodes; lo & 2002--21 & 6.78 & 19.2 & 2.83\\
onChipRam\newline DSP (Digital Signal Processors); hi & 2000--16 & 4.74 & 17.5 & 3.69\\
currentPeakPulse101000$\mu$s\newline TVS Diodes; hi & 2000--23 & 7.53 & 17.2 & 2.28\\
outputPower\newline LED Drivers Power Supplies; hi & 2010--22 & 3.33 & 16.9 & 5.08\\
framesPerSecond\newline Image Sensors, Camera; hi & 2005--23 & 9.55 & 16.5 & 1.73\\
drainToSourceVoltageVdss\newline FET, MOSFET Arrays; hi & 2002--22 & 2.13 & 16.5 & 7.73\\
responseTime\newline Photoelectric, Industrial; lo & 2000--22 & 4 & 16.5 & 4.11\\
noise10hzTo10khz\newline Voltage Reference; lo & 2000--17 & 2.31 & 16.2 & 7.03\\
currentQuiescentIq\newline Flip Flops; lo & 2001--17 & 5.71 & 16.1 & 2.81\\
currentOutput\newline Voltage Reference; hi & 2000--22 & 9.81 & 15.4 & 1.57\\
millicandelaRating\newline Panel Indicators, Pilot Lights; hi & 2001--22 & 93.8 & 13.5 & 0.144\\
currentCollectorIcMax\newline Bipolar Transistor Arrays; hi & 2002--21 & 11.4 & 13.3 & 1.17\\
currentCollectorIcMax\newline IGBT Modules; hi & 2006--22 & 3.7 & 13 & 3.5\\
voltageLoad\_max\newline Motor Drivers, Controllers; hi & 2002--22 & 1.28 & 12.7 & 9.89\\
rdsOnTyp\newline Power Distribution Switches, Load Drivers; lo & 2000--23 & 2.83 & 12.4 & 4.36\\
accessTime\newline SRAM; lo & 2001--12 & 1.3 & 12.3 & 9.44\\
onStateResistanceMax\newline Analog Switches - Special Purpose; lo & 2000--21 & 3.1 & 12.2 & 3.94\\
powerMax\newline Single Zener Diodes; hi & 2000--22 & 4.55 & 12 & 2.63\\
frequencySwitching\newline AC DC Converters, Offline Switchers; hi & 2002--22 & 9.97 & 11.5 & 1.15\\
flashSize\newline Microcontrollers, Microprocessor, FPGA Modules; hi & 2004--21 & 7.96 & 11.3 & 1.42\\
currentOutput\newline Motor Drivers, Controllers; hi & 2002--22 & 2.21 & 10.6 & 4.82\\
currentOnStateItRmsMax\newline SCRs - Modules; hi & 2002--21 & 4.91 & 10.6 & 2.16\\
outputCurrent\newline LDO Voltage Regulators; hi & 2006--22 & 2.14 & 10.2 & 4.78\\
currentOnStateItAvMax\newline SCRs - Modules; hi & 2002--21 & 5.33 & 10.1 & 1.89\\
currentPeakPulse820$\mu$s\newline Thyristors; hi & 2002--21 & 1.6 & 10 & 6.25\\
frequency\_max\newline RF Amplifiers; hi & 2002--23 & 12.2 & 9.88 & 0.812\\
temperatureCoefficient\newline Voltage Reference; lo & 2000--22 & 5.8 & 9.75 & 1.68\\
riseTime\newline MOSFETs; lo & 2010--22 & 2.9 & 9.62 & 3.32\\
3dbBandwidth\newline Analog Switches - Special Purpose; hi & 2000--21 & 1.37 & 9.37 & 6.82\\
outputCurrentChannel1\newline LED Drivers Power Supplies; hi & 2010--22 & 3.57 & 9.29 & 2.6\\
impedanceMaxZzt\newline Zener Diode Arrays; lo & 2002--18 & 5.02 & 9.22 & 1.83\\
voltageLoad\_max\newline Power Distribution Switches, Load Drivers; hi & 2000--23 & 6.55 & 9.05 & 1.38\\
powerMax\newline Bipolar Transistor Arrays; hi & 2002--21 & 8.6 & 8.81 & 1.02\\
voltageDcReverseVrMax\newline Diode Arrays; hi & 2000--23 & 11.2 & 8.54 & 0.765\\
voltageDcReverseVrMax\newline Photodiodes; hi & 2006--21 & 2.78 & 8.47 & 3.05\\
currentOutput\newline Voltage Regulators - DC DC Switching Regulators; hi & 2000--23 & 8.8 & 8.36 & 0.949\\
pdPowerDissipation\newline Zener Diodes; hi & 2002--22 & 2.05 & 8.33 & 4.06\\
powerWatts\newline AC DC Converters, Offline Switchers; hi & 2002--22 & 13.3 & 8.32 & 0.627\\
currentMax\newline LED Color Lighting; hi & 2006--23 & 1.15 & 8.23 & 7.15\\
pdPowerDissipation\newline Supervisory Circuits; hi & 2012--22 & 2.31 & 8.13 & 3.52\\
chargeCurrentMax\newline Battery Chargers; hi & 2002--22 & 4.47 & 7.51 & 1.68\\
currentMax\newline LED COBs, Engines, Modules, Strips; hi & 2010--23 & 6.5 & 7.49 & 1.15\\
supplyCurrentMax\newline SRAM; lo & 2004--18 & 3.89 & 7.39 & 1.9\\
numberOfIO\newline FPGAs (Field Programmable Gate Array); hi & 2000--23 & 2.06 & 7.35 & 3.57\\
sustainedRead\newline Solid State Drives - SSD; hi & 2013--23 & 2.95 & 7.21 & 2.45\\
sustainedWrite\newline Solid State Drives - SSD; hi & 2013--23 & 2.16 & 7.01 & 3.24\\
ifForwardCurrent\newline LED Panel Mount Indicators; hi & 2005--21 & 1 & 7 & 7\\
vrReverseVoltage\newline Rectifiers; hi & 2002--22 & 2.5 & 6.99 & 2.8\\
typicalTurnOffDelayTime\newline MOSFETs; lo & 2008--22 & 5.45 & 6.56 & 1.21\\
pdPowerDissipation\newline MOSFETs; hi & 2008--22 & 4.83 & 6.44 & 1.33\\
noiseFigure\newline RF Amplifiers; lo & 2002--23 & 3.17 & 6.29 & 1.99\\
idContinuousDrainCurrent\newline MOSFETs; hi & 2007--22 & 3.87 & 6.19 & 1.6\\
currentCollectorIcMax\newline Single, Pre-Biased Bipolar Transistors; hi & 2002--21 & 5 & 6 & 1.2\\
typicalTurnOnDelayTime\newline MOSFETs; lo & 2008--22 & 3.52 & 5.96 & 1.7\\
currentCollectorPulsedIcm\newline Single IGBTs; hi & 2004--22 & 2.51 & 5.85 & 2.33\\
inputCapacitance\newline Flip Flops; lo & 2001--17 & 1.47 & 5.73 & 3.88\\
currentAverageRectifiedIoPerDiode\newline Diode Arrays; hi & 2000--23 & 13.3 & 5.71 & 0.429\\
drainToSourceVoltageVdss\newline Single FETs, MOSFETs; hi & 2000--23 & 6.87 & 5.69 & 0.828\\
responseTime\newline Current Sensors; lo & 2002--23 & 9.83 & 5.61 & 0.57\\
currentOutputMax\newline Switches (Solid State); hi & 2002--22 & 2.61 & 5.53 & 2.12\\
frequency\_max\newline Ultrasonic Receivers, Transmitters; hi & 2011--22 & 1 & 5.46 & 5.46\\
irReverseCurrent\newline Rectifiers; lo & 2002--22 & 2.5 & 5.45 & 2.18\\
numberOfIO\newline Application Specific Microcontrollers; hi & 2004--22 & 2.07 & 5.25 & 2.54\\
voltagePeakReverseMax\newline RF Diodes; hi & 2002--22 & 3.51 & 5.21 & 1.49\\
currentCollectorIcMax\newline Single IGBTs; hi & 2002--22 & 3.32 & 5.17 & 1.56\\
dropoutVoltage\newline LDO Voltage Regulators; lo & 2006--22 & 1.72 & 5.02 & 2.93\\
switchingFrequency\newline Switching Voltage Regulators; hi & 2011--22 & 2.45 & 5.02 & 2.05\\
operatingSupplyCurrent\newline Clock Generators \& Support Products; lo & 2010--22 & 0.989 & 4.9 & 4.96\\
propagationDelayTime\newline Digital Isolators; lo & 2004--22 & 1.61 & 4.9 & 3.04\\
powerMax\newline Single IGBTs; hi & 2002--22 & 4.6 & 4.79 & 1.04\\
voltageOffState\newline SCRs; hi & 2002--22 & 2.26 & 4.78 & 2.11\\
currentSupplyMax\newline Current Sensors; lo & 2002--23 & 1.9 & 4.73 & 2.49\\
fallTime\newline MOSFETs; lo & 2010--22 & 2.07 & 4.58 & 2.21\\
frequencySwitching\newline Voltage Regulators - DC DC Switching Regulators; hi & 2000--23 & 4.14 & 4.56 & 1.1\\
maxOutputFreq\newline Clock Generators \& Support Products; hi & 2010--22 & 1.68 & 4.51 & 2.68\\
currentOutputChannel\newline Full Half-Bridge (H Bridge) Drivers; hi & 2012--22 & 9.3 & 4.39 & 0.472\\
frequencySwitching\newline DC DC Switching Controllers; hi & 2000--21 & 3.44 & 4.16 & 1.21\\
numberOfIO\newline Microcontrollers; hi & 2000--24 & 2.52 & 4.14 & 1.64\\
voltageCollectorEmitterBreakdownMax\newline Single IGBTs; hi & 2002--22 & 2 & 3.85 & 1.92\\
frequencyTransition\newline Bipolar Transistor Arrays; hi & 2002--21 & 2.58 & 3.79 & 1.47\\
currentDischargeMax820$\mu$s\newline Surge Protection Devices (SPDs); hi & 2011--23 & 2.9 & 3.68 & 1.27\\
frequencyTransition\newline Single Bipolar Transistors; hi & 2000--22 & 2.07 & 3.64 & 1.76\\
voltageForwardVfTyp\newline LED Emitters - Infrared, UV, Visible; lo & 2002--22 & 1.2 & 3.63 & 3.02\\
currentOnStateItRmsMax\newline TRIACs; hi & 2002--21 & 4.38 & 3.54 & 0.808\\
activeReadCurrentMax\newline NOR Flash; lo & 2002--21 & 1.77 & 3.33 & 1.88\\
voltageOffState\newline SCRs - Modules; hi & 2002--21 & 1.29 & 3.25 & 2.52\\
pdPowerDissipation\newline Transient Voltage Suppression Diodes - TVS; hi & 2002--22 & 1.39 & 3.21 & 2.32\\
numberOfBits\newline Digital to Analog Converters (DAC); hi & 2000--22 & 1.33 & 3.19 & 2.39\\
clockRate\newline DSP (Digital Signal Processors); hi & 2000--16 & 2.97 & 3.17 & 1.06\\
powerMax\newline Bipolar Transistor Arrays, Pre-Biased; hi & 2002--21 & 1.03 & 3.08 & 3\\
voltageCollectorEmitterBreakdownMax\newline Bipolar Transistor Arrays; hi & 2002--21 & 2.08 & 3.04 & 1.46\\
numberOfIO\newline System On Chip (SoC); hi & 2011--23 & 2.09 & 3.02 & 1.45\\
voltageCollectorEmitterBreakdownMax\newline IGBT Modules; hi & 2006--22 & 1.42 & 3 & 2.12\\
currentOutputChannel\newline Transistor, Photovoltaic Output Optoisolators; hi & 2000--22 & 3 & 3 & 1\\
currentOutputTyp\newline Comparators; hi & 2000--17 & 1.62 & 2.84 & 1.75\\
currentQuiescentMax\newline Gates and Inverters; lo & 2000--22 & 1.83 & 2.83 & 1.54\\
vfForwardVoltage\newline Transient Voltage Suppression Diodes - TVS; lo & 2002--22 & 1.17 & 2.79 & 2.39\\
powerMax\newline Single, Pre-Biased Bipolar Transistors; hi & 2002--21 & 2.55 & 2.69 & 1.06\\
lumensWattCurrentTest\newline LED Color Lighting; hi & 2006--23 & 1.45 & 2.69 & 1.86\\
currentMax\newline LED White Lighting; hi & 2006--23 & 1.22 & 2.62 & 2.16\\
pdPowerDissipation\newline Digital Isolators; hi & 2008--22 & 1.69 & 2.61 & 1.55\\
voltageBreakdown\newline AC DC Converters, Offline Switchers; hi & 2002--23 & 1.04 & 2.39 & 2.29\\
dropoutVoltageMax\newline LDO Voltage Regulators; lo & 2012--22 & 2.18 & 2.37 & 1.09\\
voltageBreakover\newline Thyristors; hi & 2002--22 & 1.71 & 2.29 & 1.34\\
capacitance\newline Thyristors; lo & 2002--21 & 2.16 & 2.28 & 1.06\\
voltageIsolation\newline Power Driver Modules; hi & 2004--22 & 1.03 & 2.24 & 2.18\\
currentPeakPulse101000$\mu$s\newline Thyristors; hi & 2002--22 & 1.25 & 2.24 & 1.79\\
voltageOffState\newline Thyristors; hi & 2002--22 & 1.68 & 2.17 & 1.29\\
vfForwardVoltage\newline Rectifiers; lo & 2002--22 & 1.23 & 2.06 & 1.67\\
gain$^{b}$\newline RF Amplifiers; hi & 2002--23 & 1.61 & 2.01 & 1.25\\
currentDischargeNom820$\mu$s\newline Surge Protection Devices (SPDs); hi & 2011--23 & 1 & 2 & 2\\
resolutionBits\newline ADCs/DACs - Special Purpose; hi & 2001--22 & 1 & 2 & 2\\
loadVoltageRating\_max\newline Solid State Relays - Industrial Mount; hi & 2007--22 & 2.01 & 1.86 & 0.922\\
currentDcForwardIfMax\newline Transistor, Photovoltaic Output Optoisolators; hi & 2000--22 & 1.56 & 1.83 & 1.17\\
maximumClockFrequency\newline SRAM; hi & 2001--17 & 1.5 & 1.76 & 1.18\\
numberOfIO\newline I/O Expanders; hi & 2002--21 & 2 & 1.66 & 0.831\\
voltageCollectorEmitterBreakdownMax\newline Single, Pre-Biased Bipolar Transistors; hi & 2002--21 & 1 & 1.6 & 1.6\\
voltageCollectorEmitterBreakdownMax\newline Bipolar Transistor Arrays, Pre-Biased; hi & 2002--21 & 1 & 1.6 & 1.6\\
psrrRippleRejectionTyp\newline LDO Voltage Regulators; hi & 2007--22 & 1.02 & 1.6 & 1.57\\
voltageOffState\newline TRIACs; hi & 2002--21 & 1.33 & 1.51 & 1.13\\
voltageRegulationAccuracy\newline LDO Voltage Regulators; lo & 2006--22 & 2 & 1.44 & 0.721\\
pulseWidth\newline Digital Isolators; lo & 2006--22 & 5.91 & 1.39 & 0.235\\
frequencySwitching\newline Voltage Regulators - Linear + Switching; hi & 2007--22 & 1.1 & 1.38 & 1.26\\
currentDcForwardIfMax\newline Triac, SCR Output Optoisolators; hi & 2002--17 & 1.45 & 1.36 & 0.938\\
vceSaturationMax\newline Transistor, Photovoltaic Output Optoisolators; lo & 2000--22 & 1.99 & 1.28 & 0.644\\
isolationVoltage\newline Digital Isolators; hi & 2004--22 & 1.5 & 1.21 & 0.805\\
voltageIsolation\newline Transistor, Photovoltaic Output Optoisolators; hi & 2000--22 & 1.07 & 1.06 & 0.994\\
\end{longtable}
\noindent\footnotesize $^{a}$ Unresolved pin-specific current measurement. $^{b}$ Gain recorded on a dB scale.\par
\normalsize

\clearpage

\section{Measurement and output provenance}\label{app:provenance}
Table~\ref{tab:audit} summarizes the implementation audit. The empirical figures use the reported analytical outputs. Fig.~\ref{fig:variables} combines an observed source trajectory with schematic variable definitions, while Figs.~\ref{fig:population}--\ref{fig:forecast} present the empirical summaries. The p99 frontier construction is defined in Section~\ref{sec:frontier}.

The analysis scripts and input data for reproducing the manuscript’s figures and empirical tables are available in the Git repository identified in the Code availability statement. The inputs comprise the analysis catalog, capability-selection table, and saved capability-level shoulder estimates.

The anchor measurements use the historical product database and the corresponding analysis scripts in the repository. The revised 204-capability logistic comparison gives mean paired AUC difference $-0.04167$ and weighted difference $-0.03610$, compared with $-0.04105$ and $-0.03539$ in the original panel. Fig.~\ref{fig:forecast}, its calibration summaries and the boosting sensitivity were regenerated on the retained panel. The full 206-capability shoulder bootstrap and null procedure was rerun in its original feature order and reproduced the saved output exactly; the reported aggregates restrict those verified estimates to the retained set. Figs.~\ref{fig:population} and \ref{fig:patterns} and Appendices~\ref{app:leadpanel} and~\ref{app:trajectories} identify their respective samples. Timing and stratification were rerun on the same retained panel, replacing the historical diagnostics. The repository reconstructs the retained forecast panel from the analysis catalog and uses it to regenerate the forecast and trajectory figures. The associated outputs include annual diagnostics, paired predictions, calibration-bin counts, and annual median trajectories; the underlying primary forecast and calibration results are unchanged.

The interarrival and stratification comparisons were rerun on the retained panel using a wrapper around the original analysis scripts. The wrapper preserves the permutation counts and estimators, fixes tied-depth ordering for reproducibility, and saves scored predictions and complete diagnostic outputs. Capability-coupling and pooled stall comparisons were recomputed using the original analysis scripts, reproducing the reported values. The repository pins its Python dependencies in \texttt{requirements-lock.txt}.

\begin{table}[htbp]
\centering\small
\caption{Implementation audit: resolved definitions and remaining provenance questions}\label{tab:audit}
\begin{tabularx}{\textwidth}{P{0.27\textwidth}Y}
\toprule
Analysis & Verified implementation or remaining question \\\midrule
Historical data & Acquisition dates, precise meaning of the release-year field, duplicates and variants, source-specific unit checks. \\
p95 gap illustrations & Gaussian smoothing of log$_{10}$ annual medians, bandwidth 1.2 observation positions; two-sided, edge-normalized weights. \\
Four history variables & Changes compare usable positions $i-1$ and $i-4$; age is calendar time since the last earlier advance, missing before a first event. \\
Predictive pipeline & Training-fold median imputation and standardization; logistic lbfgs, $C=1$, 1,000 iterations. All-missing columns drop. Boosting and calibration settings are stated in the methods. \\
Timing diagnostics & Logistic age slope ($C=10^6$); mean within-capability CV; 2,000 shuffles, seed 0; event-count-preserving full-window and eight-year-block references; tied values included in empirical tails. \\
Stratification and cases & Retained-panel predictions regenerated; 160 eligible capabilities; 5,000 depth shuffles and 2,000 per-cut maximum-statistic shuffles. Coupling uses separately detrended log medians and ties-inclusive quartile groups; case and pooled stall values were reproduced. \\
\bottomrule
\end{tabularx}
\end{table}

\subsection{Excluded interpretations of earlier exploratory analyses}
The original attempts-space producer records complete interrecord waits and capability gains. These are not the attempt-indexed age and residual waiting time analyzed by \citet{yinwang2026}; its gain/wait association is not a reproduction or falsification of that relationship. An earlier threshold-heterogeneity simulation includes current-step gains among predictors of current replacement and pools buyer-year rows rather than a superposed stream. Its AUC pattern is not retained as prospective forecasting evidence or as a proof of an aggregation theorem. The restricted analytical timing example in Appendix~\ref{app:timingtheory} is explicitly separate.

\clearpage

\section{Empirical trajectories across the anchor capabilities}\label{app:trajectories}
\begin{figure}[p]\centering\includegraphics[width=\textwidth]{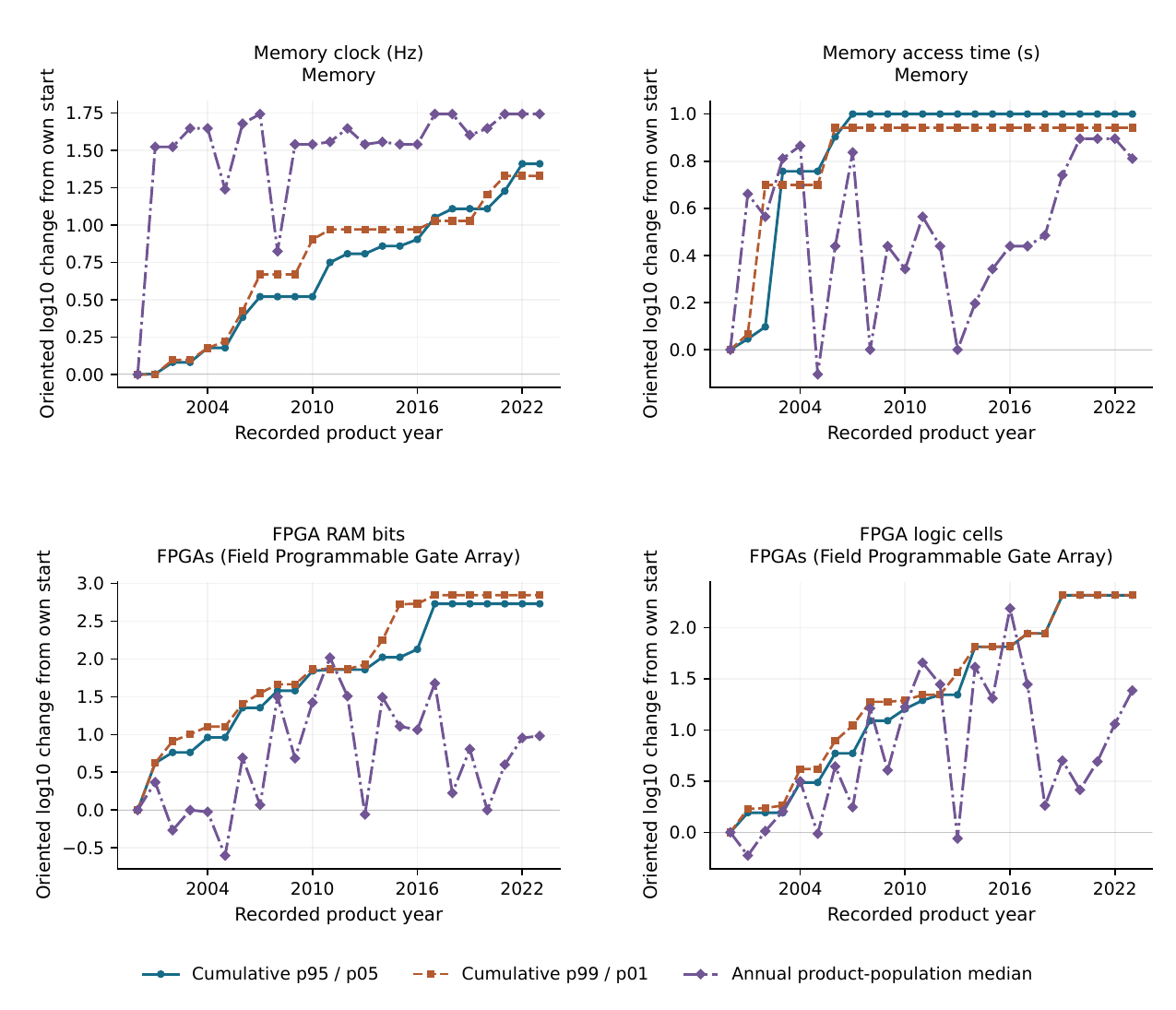}\caption{Anchor trajectories, group 1. Solid circles: cumulative p95/p05; dashed squares: cumulative p99/p01; dash-dot diamonds: annual product-population median. Each series is normalized to its own start, so the lines compare relative changes rather than absolute specification levels}\label{fig:trajectories01}\end{figure}

\begin{figure}[p]\centering\includegraphics[width=\textwidth]{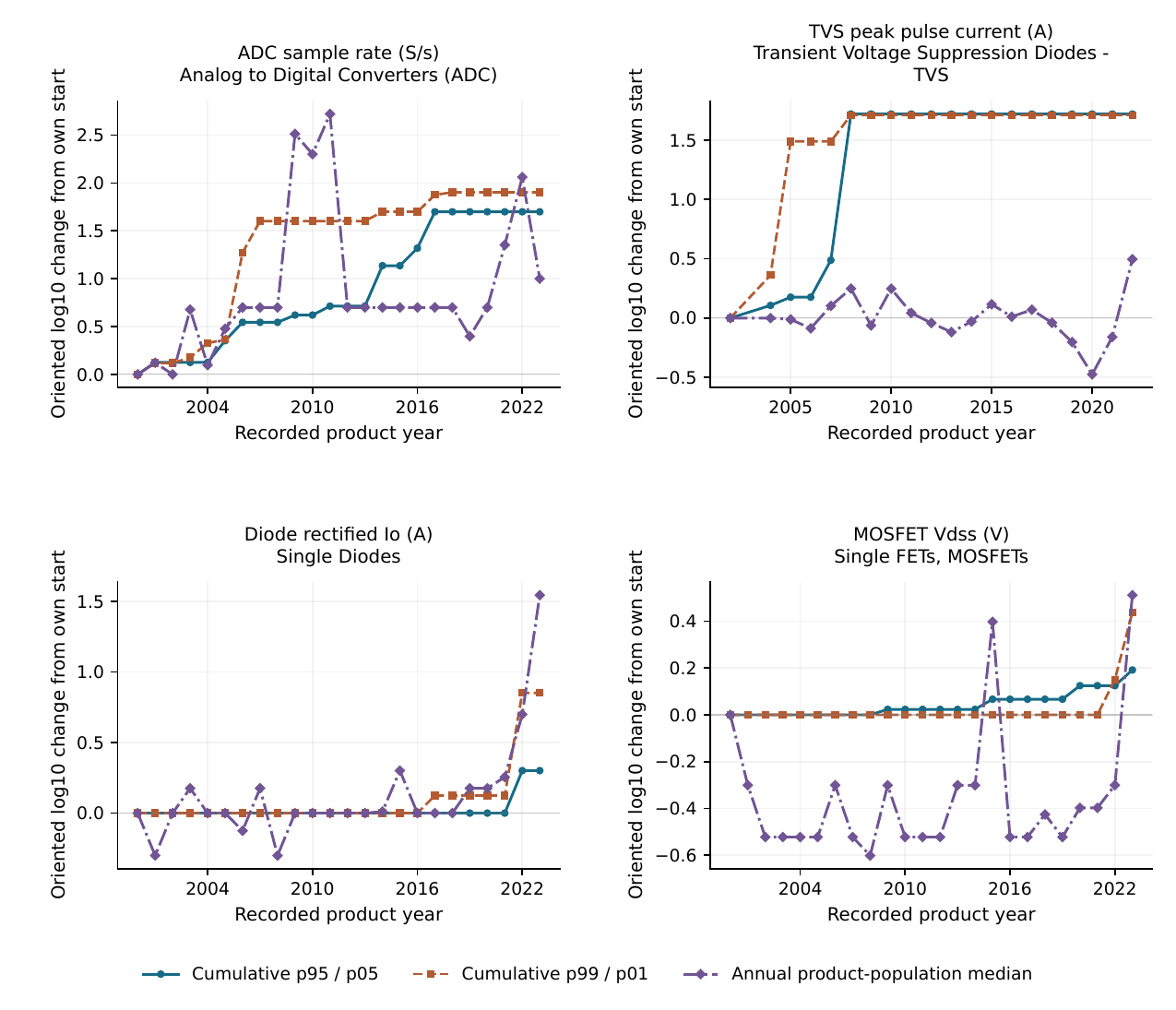}\caption{Anchor trajectories, group 2. Solid circles: cumulative p95/p05; dashed squares: cumulative p99/p01; dash-dot diamonds: annual product-population median. Each series is normalized to its own start, so the lines compare relative changes rather than absolute specification levels}\label{fig:trajectories02}\end{figure}

\begin{figure}[p]\centering\includegraphics[width=\textwidth]{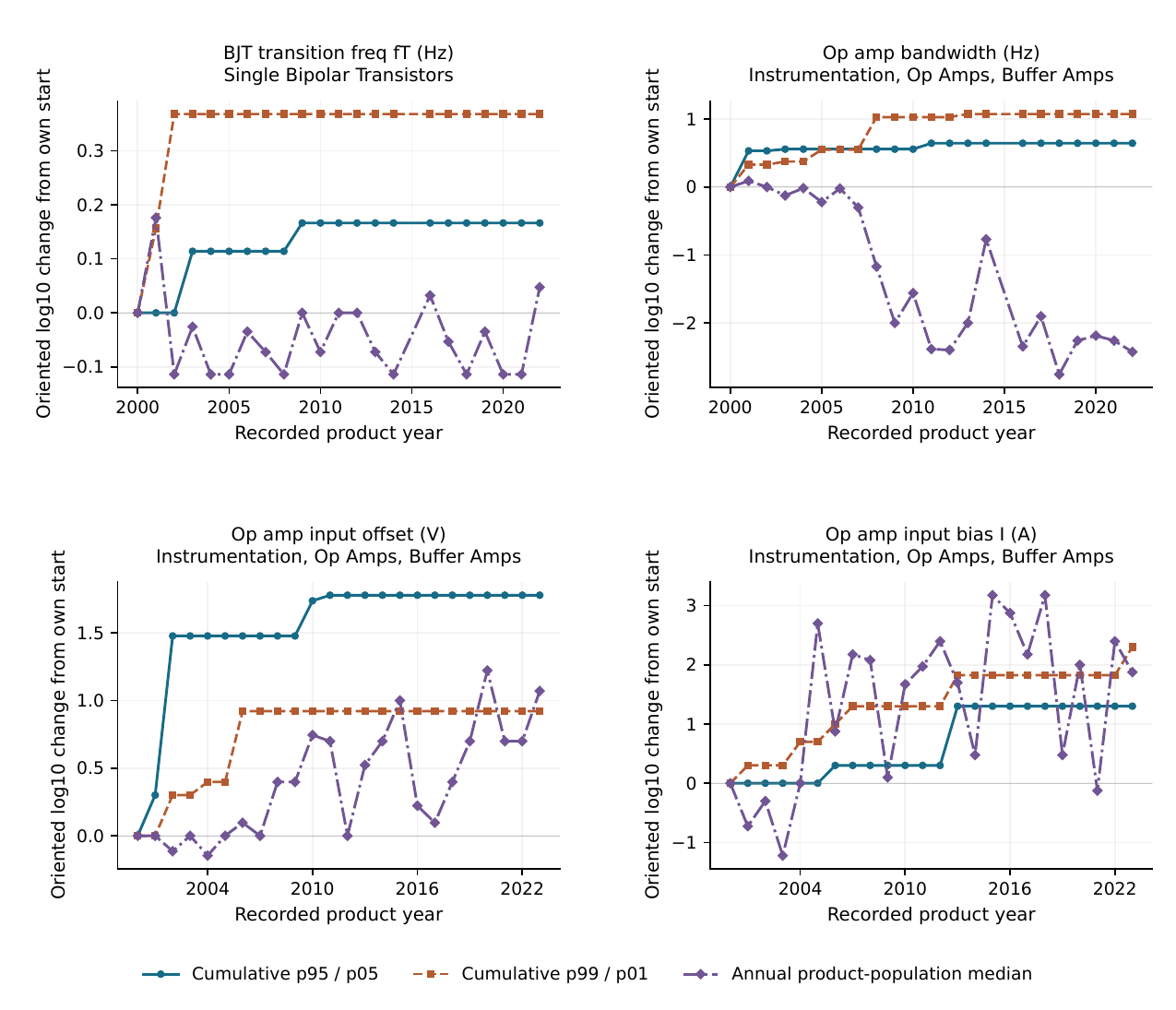}\caption{Anchor trajectories, group 3. Solid circles: cumulative p95/p05; dashed squares: cumulative p99/p01; dash-dot diamonds: annual product-population median. Each series is normalized to its own start, so the lines compare relative changes rather than absolute specification levels}\label{fig:trajectories03}\end{figure}

\begin{figure}[p]\centering\includegraphics[width=\textwidth]{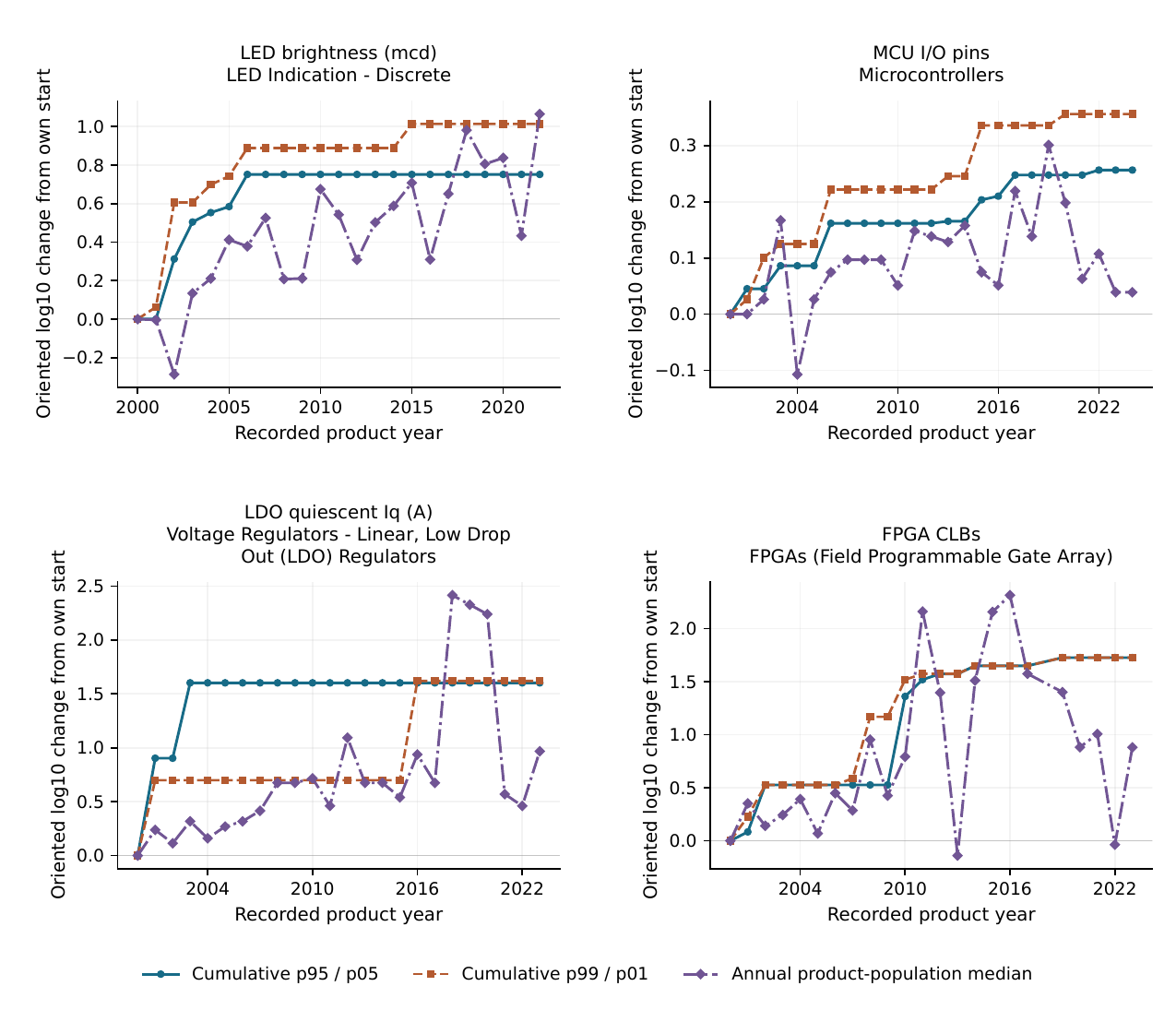}\caption{Anchor trajectories, group 4. Solid circles: cumulative p95/p05; dashed squares: cumulative p99/p01; dash-dot diamonds: annual product-population median. Each series is normalized to its own start, so the lines compare relative changes rather than absolute specification levels}\label{fig:trajectories04}\end{figure}

\begin{figure}[p]\centering\includegraphics[width=\textwidth]{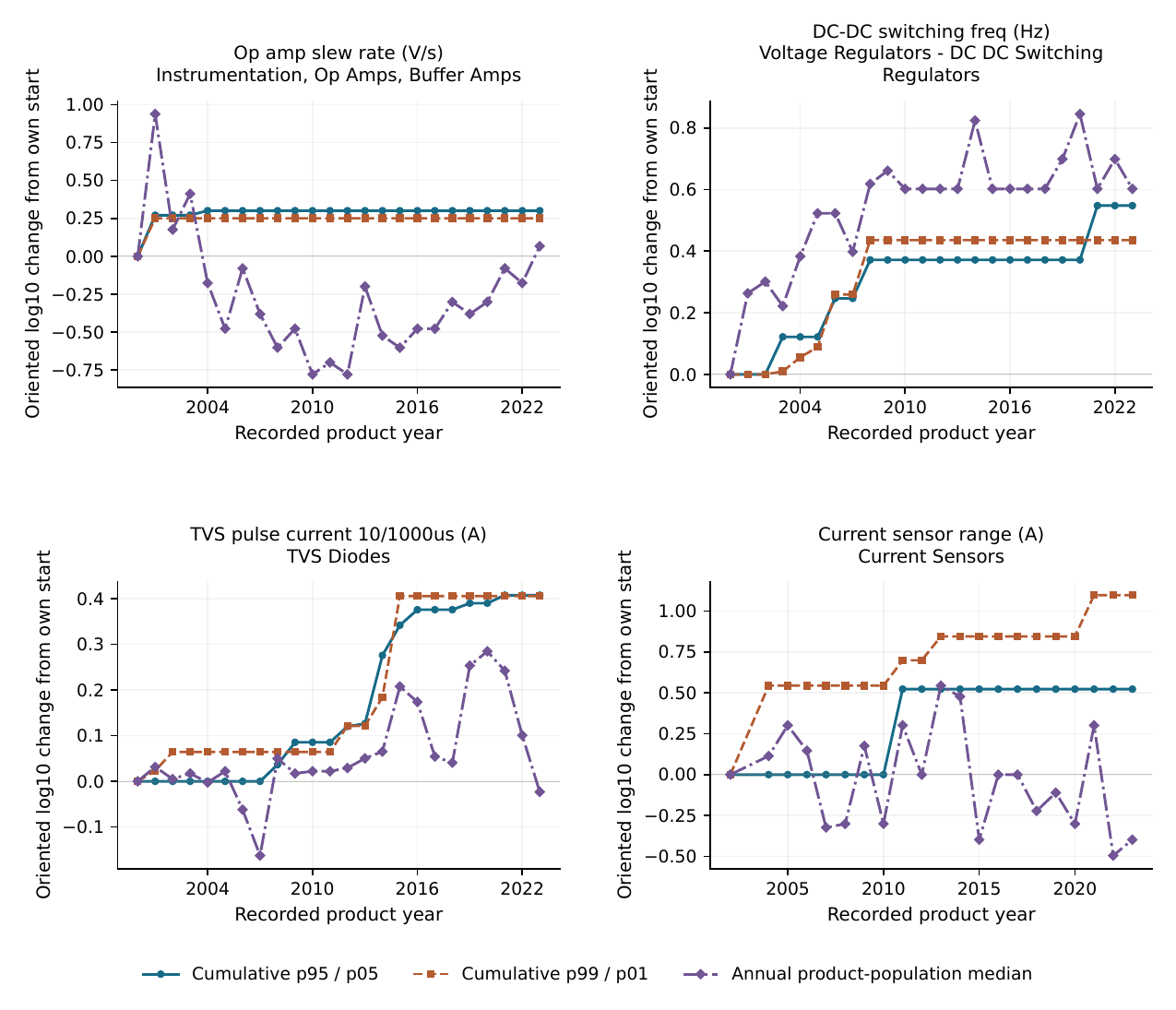}\caption{Anchor trajectories, group 5. Solid circles: cumulative p95/p05; dashed squares: cumulative p99/p01; dash-dot diamonds: annual product-population median. Each series is normalized to its own start, so the lines compare relative changes rather than absolute specification levels}\label{fig:trajectories05}\end{figure}

\begin{figure}[p]\centering\includegraphics[width=\textwidth]{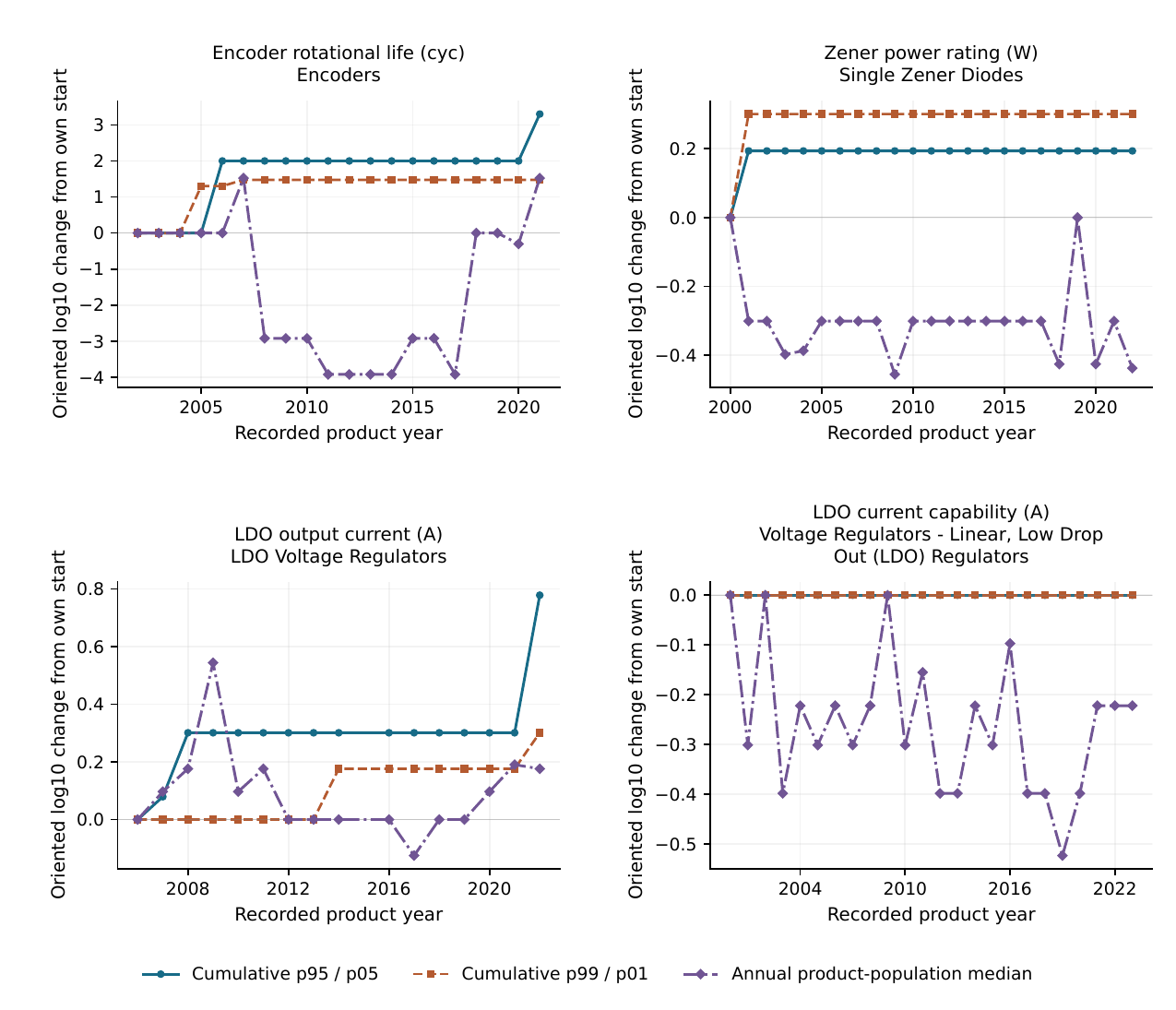}\caption{Anchor trajectories, group 6. Solid circles: cumulative p95/p05; dashed squares: cumulative p99/p01; dash-dot diamonds: annual product-population median. Each series is normalized to its own start, so the lines compare relative changes rather than absolute specification levels}\label{fig:trajectories06}\end{figure}

\begin{figure}[p]\centering\includegraphics[width=\textwidth]{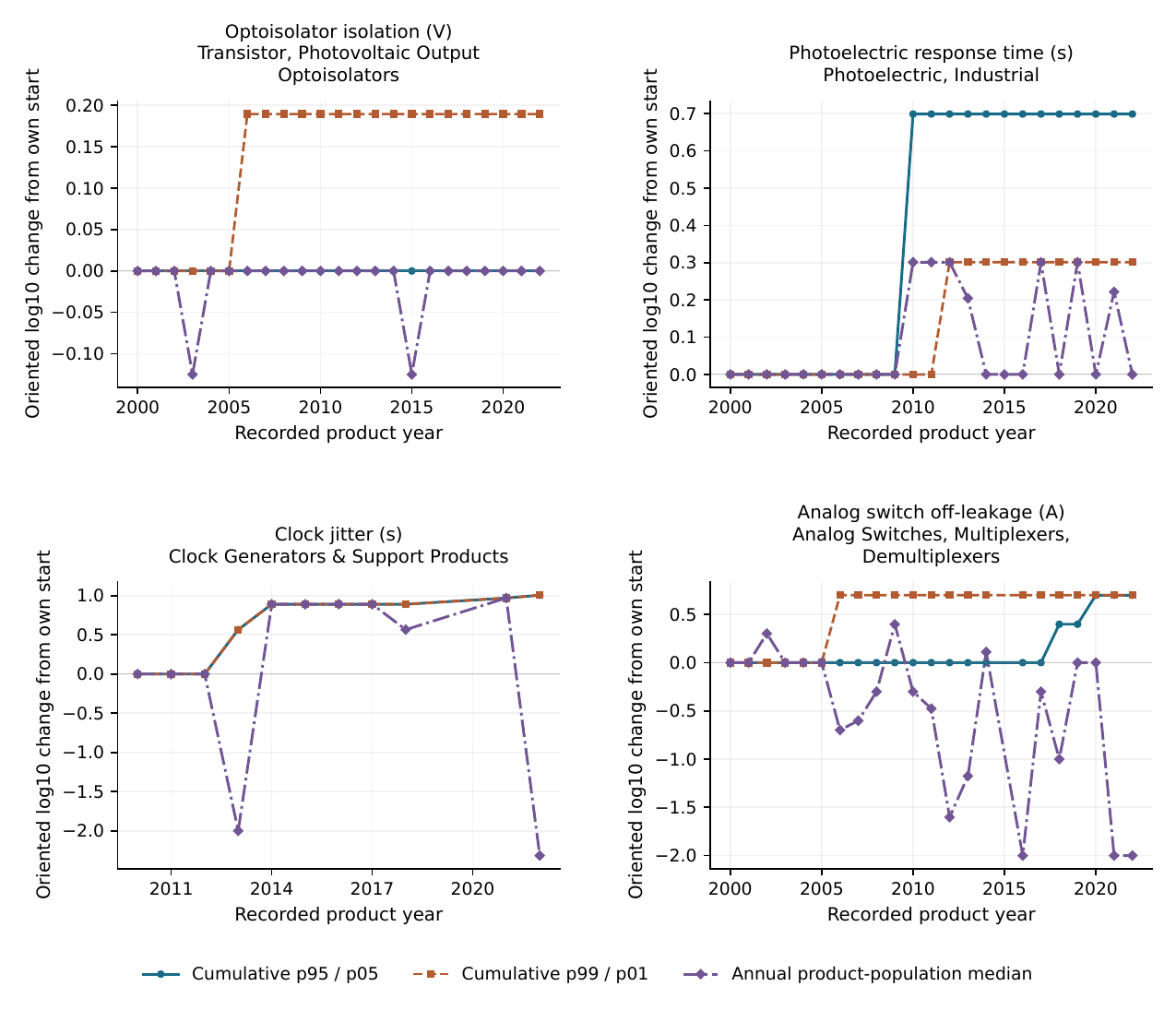}\caption{Anchor trajectories, group 7. Solid circles: cumulative p95/p05; dashed squares: cumulative p99/p01; dash-dot diamonds: annual product-population median. Each series is normalized to its own start, so the lines compare relative changes rather than absolute specification levels}\label{fig:trajectories07}\end{figure}

Figs.~\ref{fig:trajectories01}--\ref{fig:trajectories07} display all 28 original anchor capabilities with a ratio-scale
interpretation, spanning memory, logic, analog, power, sensing, and optoelectronics.
The original RF-amplifier gain anchor is omitted here because gain is recorded in
dB; taking a log-ratio of dB values does not measure proportional physical gain.
No selection is made on visual punctuation, lead-ratio magnitude, or forecasting
performance. Thus both concentrated advances and relatively flat or gradual
trajectories are visible. These anchors were originally selected manually and
are not a representative random sample of the commercial product population.

Each panel shows direction-adjusted cumulative annual p95 and p99 frontiers
(p05 and p01 for lower-is-better attributes) together with the annual commercial
product-population median. The median uses the same positive, non-null records
and usable years as the frontier and is neither smoothed nor cumulated. Each
of the three series is normalized to its own first usable value: the plotted
quantity is $d_j\log_{10}(X_{jt}/X_{j,t_0})$, where $X$ is the corresponding
frontier or median. A rise of one unit represents tenfold improvement relative
to that series' own start. The panels compare relative changes over time;
vertical distances and crossings do not measure absolute frontier--median gaps.
Unlike cumulative frontiers, the annual median can rise or fall as the observed
product population changes. Points denote usable source years;
lines join observations and do not impute intervening years. Catalog years have
not been verified as introduction dates. A cumulative-best construction is
monotone by definition: staircase appearance alone is not evidence for a
punctuated-equilibrium mechanism. Quantiles require at least 40 positive listings
per usable year and at least eight usable years per capability.


\clearpage

\section{Economic interpretation benchmarks}\label{app:interpretation}
\subsection{Finite-horizon adoption comparison}
Under the exponential incremental-service-flow assumptions of equation~\eqref{eq:threshold}, the finite-horizon requirement is
\begin{equation}\label{eq:finite}
 \Delta q^*(T)=\frac{(\rho+\delta)\mathcal C}
 {w[1-\mathrm{e}^{-(\rho+\delta)T}]}.
\end{equation}
Its infinite-horizon limit is equation~\eqref{eq:threshold}. For $q>0$, the ratio $\Delta q^*/q=(\rho+\delta)\mathcal C/(wq)$ is dimensionless, not a duration. With nonlinear service values, the full incremental benefit in equation~\eqref{eq:wtp} must replace a constant-$w$ extrapolation.

For illustration, a fixed redesign cost $\mathcal F=\$100,000$, incremental unit cost $c=0$, and annualization rate $\rho+\delta=0.12$ imply unit switching costs of \$10, \$1, and \$0.10 at production volumes of 10,000, 100,000, and 1,000,000. The corresponding required annual incremental benefits are \$1.20, \$0.12, and \$0.012 per unit. None of those volumes or service values is measured in the product data. A required percentage performance gain cannot be recovered without a benefit function.

For a second hypothetical example, incremental annual net service of \$20,000 over three years at a 10\% continuous discount rate has value $20,000(1-\mathrm{e}^{-0.3})/0.1\simeq\$51,836$. This slightly exceeds a \$50,000 net switching cost. At \$15,000 annual incremental benefit, the corresponding value is approximately \$38,877. Both compare incremental service with cost, not the total earnings of a new asset with an assumed performance ratio. Neither identifies an optimal replacement date.

\subsection{Threshold accumulation and the event being modeled}\label{app:timingtheory}
Suppose equal technical opportunities of size $a_q$ arrive at Poisson rate $\lambda$ and a buyer replaces after cumulative improvement reaches $\theta$. Then $k=\lceil\theta/a_q\rceil$ opportunities are required, with waiting-time survival
\begin{equation}\label{eq:erlang}
 \Prob(T>t)=\mathrm{e}^{-\lambda t}\sum_{m=0}^{k-1}\frac{(\lambda t)^m}{m!},\quad
 \E[T]=\frac{k}{\lambda},\quad
 \operatorname{Var}(T)=\frac{k}{\lambda^2}.
\end{equation}
The coefficient of variation is $1/\sqrt{k}$ and the hazard increases for $k>1$ in this restricted example. Random increments, changing thresholds, and forward-looking decisions need not preserve these properties. This model concerns \emph{buyer replacement}, whereas the empirical intervals concern \emph{frontier advances}. It is not fitted to the 109 gaps in the retained timing panel.

Superposition results require suitable independence, rarity and intensity conditions \citep{cinlar1972superposition}. Buyers exposed to common standards or demand shocks need not satisfy them. Pooling waiting-time observations across capabilities is not the same operation as superposing their event streams. Neither the Erlang example nor a superposition theorem predicts the sign of the empirical AUC difference.

\subsection{Asset value and revaluation}\label{app:assetvalue}
A benchmark value of an installed asset is
\begin{equation}\label{eq:assetvalue}
 V_t=\E_t\left[\int_0^{T_t}\mathrm{e}^{-\rho s}m(t,s)\,\mathrm{d}s
 +\mathrm{e}^{-\rho T_t}S_{t+T_t}\right],
\end{equation}
where $m(t,s)$ is expected net service flow, $T_t$ remaining life, and $S$ terminal value. This values an asset, whereas equation~\eqref{eq:wtp} values an incremental replacement benefit. Consistent treatment of incumbent income and terminal value is required to avoid double counting.

In a restricted example with constant service quantity $\bar q$, zero operating costs, infinite life, and deterministic service price $p(t+s)=p(t)\mathrm{e}^{-gs}$ with $\rho+g>0$,
\begin{equation}
 V_t=\frac{\bar q p(t)}{\rho+g},\qquad -\dot V_t/V_t=g.
\end{equation}
Value changes here reflect the assumed service-price path, not physical deterioration or a measured age-price depreciation schedule \citep{oecd2009capital}. Changes in operating cost, utilization, physical age and remaining life alter the comparison. A revision to future net rents can raise value, but a frontier advance alone does not determine whether installed equipment appreciates or depreciates. The product database contains neither the service nor asset prices needed to test this extension.

\clearpage

\section{Contextual information and representation}\label{app:information}
For product-feature history $X$, an independently measured environment $R$, and a discrete future outcome $Y$,
\begin{equation}\label{eq:information}
 H(Y\mid X)-H(Y\mid X,R)=I(Y;R\mid X)\geq0.
\end{equation}
This standard identity describes a potential average reduction in uncertainty \citep{polyanskiywu2016}. It does not establish that the particular proposed environment has nonzero conditional information, or that a finite fitted model can exploit it. For a language-model output $Z$ constructed from $X$ and external knowledge $E$, without additional access to $Y$, data processing gives $I(Y;Z)\leq I(Y;X,E)$. Historical knowledge retained by a pretrained model belongs in $E$. Transforming $X$ can improve its representation without adding information beyond $X$; an explicit regime label is not mathematically necessary when a model already represents the relevant distinctions. Neither the current AUC nor its change estimates equation~\eqref{eq:information}.

\clearpage

\section{Exploratory language-model pilots}\label{app:llmpilots}
The original exploratory elicitation uses 162 feature-years and 30 advances across FPGA and LDO capabilities. Pooled AUC is approximately 0.535 when the year is omitted and 0.605 when it is provided. This retrospective experiment evaluates elicited scores for historical events. Prompt anchoring---sensitivity to how contextual information is framed in the prompt---and batching affected scores, and omission of the year does not rule out historical knowledge.

A minimal-pair ablation uses 12 base cases, two passes, and nine changed facts, including a packaging placebo. Responses are repeatable, but the vector of responses to eight substantive changes has rank correlation approximately $-0.07$ with their observed univariate associations with advances. The packaging manipulation changes the elicited score by approximately nine points. The mismatch between elicited responses and observed associations raises a concern about instrument validity; causal interpretation requires independent evidence.

A separate 60-query attribution exercise pairs 30 actual advances with 30 non-advance feature-years. The model affirms an external event for 15 of 19 real FPGA advances and one of 11 real LDO advances. A score combining affirmation, specificity and confidence yields AUC approximately 0.66 for FPGA and 0.49 for LDO. These scores concern the elicited responses; independent factual verification remains unfinished.

\clearpage
\FloatBarrier
\bibliographystyle{plainnat}
\bibliography{ref}

@article{anderson1990technological,
  author  = {Anderson, Philip and Tushman, Michael L.},
  title   = {Technological Discontinuities and Dominant Designs: A Cyclical Model of Technological Change},
  journal = {Administrative Science Quarterly},
  year    = {1990},
  volume  = {35},
  number  = {4},
  pages   = {604--633},
  doi     = {10.2307/2393511}
}

@incollection{cinlar1972superposition,
  author    = {\c{C}inlar, Erhan},
  title     = {Superposition of point processes},
  booktitle = {Stochastic Point Processes: Statistical Analysis, Theory, and Applications},
  editor    = {Lewis, Peter A. W.},
  publisher = {Wiley},
  address   = {New York},
  pages     = {549--606},
  year      = {1972}
}

@book{dixitpindyck1994,
 author = {Dixit, Avinash K. and Pindyck, Robert S.},
 title = {Investment under Uncertainty},
 publisher = {Princeton University Press},
 address = {Princeton, NJ},
 year = {1994},
 doi = {10.2307/j.ctt7sncv}
}

@article{dopfer2004micro,
  author  = {Dopfer, Kurt and Foster, John and Potts, Jason},
  title   = {Micro--Meso--Macro},
  journal = {Journal of Evolutionary Economics},
  year    = {2004},
  volume  = {14},
  pages   = {263--279},
  doi     = {10.1007/s00191-004-0193-0}
}

@article{dosi1982,
  author  = {Dosi, Giovanni},
  title   = {Technological Paradigms and Technological Trajectories: A Suggested Interpretation of the Determinants and Directions of Technical Change},
  journal = {Research Policy},
  year    = {1982},
  volume  = {11},
  number  = {3},
  pages   = {147--162},
  doi     = {10.1016/0048-7333(82)90016-6}
}

@article{dosi1994introduction,
  author  = {Dosi, Giovanni and Nelson, Richard R.},
  title   = {An Introduction to Evolutionary Theories in Economics},
  journal = {Journal of Evolutionary Economics},
  year    = {1994},
  volume  = {4},
  pages   = {153--172},
  doi     = {10.1007/BF01236366}
}

@article{fontana2009product,
  author  = {Fontana, Roberto and Nuvolari, Alessandro and Saviotti, Pier Paolo},
  title   = {Introduction to the Journal of Evolutionary Economics Special Issue: The Product Characteristics Approach to Innovation Studies},
  journal = {Journal of Evolutionary Economics},
  year    = {2009},
  volume  = {19},
  pages   = {463--469},
  doi     = {10.1007/s00191-009-0139-7}
}

@article{frenken1999variety,
  author  = {Frenken, Koen and Saviotti, Pier Paolo and Trommetter, Michel},
  title   = {Variety and Niche Creation in Aircraft, Helicopters, Motorcycles and Microcomputers},
  journal = {Research Policy},
  year    = {1999},
  volume  = {28},
  number  = {5},
  pages   = {469--488},
  doi     = {10.1016/S0048-7333(99)00008-6},
  url     = {https://doi.org/10.1016/S0048-7333(99)00008-6}
}

@article{gersick1991revolutionary,
  author  = {Gersick, Connie J. G.},
  title   = {Revolutionary Change Theories: A Multilevel Exploration of the Punctuated Equilibrium Paradigm},
  journal = {Academy of Management Review},
  year    = {1991},
  volume  = {16},
  number  = {1},
  pages   = {10--36},
  doi     = {10.5465/amr.1991.4278988}
}

@article{gould1977punctuated,
  author  = {Gould, Stephen Jay and Eldredge, Niles},
  title   = {Punctuated Equilibria: The Tempo and Mode of Evolution Reconsidered},
  journal = {Paleobiology},
  year    = {1977},
  volume  = {3},
  number  = {2},
  pages   = {115--151},
  doi     = {10.1017/S0094837300005224}
}

@article{he2026benchmark,
  author        = {He, Youwei and Lee, Jeong-Dong},
  doi           = {10.1016/j.techfore.2026.124702},
  journal       = {Technological Forecasting and Social Change},
  pages         = {124702},
  title         = {Forecasting Benchmark Product Emergence via Technological Phylogeny and Strategic Features},
  volume        = {229},
  year          = {2026}
}

@article{henderson1990architectural,
  author  = {Henderson, Rebecca M. and Clark, Kim B.},
  title   = {Architectural Innovation: The Reconfiguration of Existing Product Technologies and the Failure of Established Firms},
  journal = {Administrative Science Quarterly},
  year    = {1990},
  volume  = {35},
  number  = {1},
  pages   = {9--30},
  doi     = {10.2307/2393549}
}

@article{jeong2024where,
  author  = {Jeong, Dawoon and Lee, Jeong-Dong},
  title   = {Where and How Does a Product Evolve? Product Innovation Pattern in Product Lineage},
  journal = {Technovation},
  year    = {2024},
  volume  = {131},
  pages   = {102958},
  doi     = {10.1016/j.technovation.2024.102958}
}

@article{jeong2025mapping,
  author  = {Jeong, Dawoon and Lee, Jeong-Dong},
  title   = {Mapping Product Development Trajectories: Product Citation Network},
  journal = {Scientometrics},
  year    = {2025},
  volume  = {130},
  number  = {8},
  pages   = {4667--4696},
  doi     = {10.1007/s11192-025-05395-y}
}

@article{klemperer1995switching,
 author = {Klemperer, Paul},
 title = {Competition when Consumers Have Switching Costs: An Overview with Applications to Industrial Organization, Macroeconomics, and International Trade},
 journal = {The Review of Economic Studies},
 year = {1995},
 volume = {62},
 number = {4},
 pages = {515--539},
 doi = {10.2307/2298075}
}

@article{kosonocky2024mining,
  author        = {Kosonocky, Clayton W. and Wilke, Claus O. and Marcotte, Edward M. and
Ellington, Andrew D.},
  doi           = {10.1039/d4dd00011k},
  journal       = {Digital Discovery},
  pages         = {1150--1159},
  title         = {Mining Patents with Large Language Models Elucidates the Chemical
Function Landscape},
  number        = {6},
  volume        = {3},
  year          = {2024}
}

@article{lee2004estimating,
  author  = {Lee, Jeong-Dong and Repkine, Alexandre and Hwang, Seog-Won and Kim, Tai-Yoo},
  title   = {Estimating consumers' willingness to pay for the individual quality attributes with {DEA}},
  journal = {Journal of the Operational Research Society},
  volume  = {55},
  number  = {10},
  pages   = {1064--1070},
  year    = {2004},
  doi     = {10.1057/palgrave.jors.2601797}
}

@article{lee2024mapping,
  author  = {Lee, Jeong-Dong and Jeong, Dawoon and Jung, Euy-Young and Kim, Yunyoung and Kim, Jiyong and He, Youwei and Choi, Sungjun},
  title   = {Mapping the Evolutionary Pattern of Mobile Products: A Phylogenetic Approach},
  journal = {IEEE Transactions on Engineering Management},
  year    = {2024},
  volume  = {71},
  pages   = {4776--4790},
  doi     = {10.1109/TEM.2022.3214489}
}

@article{loch1999punctuated,
  author  = {Loch, Christoph H. and Huberman, Bernardo A.},
  title   = {A Punctuated-Equilibrium Model of Technology Diffusion},
  journal = {Management Science},
  year    = {1999},
  volume  = {45},
  number  = {2},
  pages   = {160--177},
  doi     = {10.1287/mnsc.45.2.160}
}

@article{march1991exploration,
  author  = {March, James G.},
  title   = {Exploration and Exploitation in Organizational Learning},
  journal = {Organization Science},
  year    = {1991},
  volume  = {2},
  number  = {1},
  pages   = {71--87},
  doi     = {10.1287/orsc.2.1.71}
}

@article{metcalfe1994competition,
  author  = {Metcalfe, J. Stanley},
  title   = {Competition, Fisher's Principle and Increasing Returns in the Selection Process},
  journal = {Journal of Evolutionary Economics},
  year    = {1994},
  volume  = {4},
  pages   = {327--346},
  doi     = {10.1007/BF01236409}
}

@book{nelsonwinter1982,
  author    = {Nelson, Richard R. and Winter, Sidney G.},
  title     = {An Evolutionary Theory of Economic Change},
  year      = {1982},
  publisher = {Harvard University Press},
  address   = {Cambridge, MA}
}

@book{oecd2009capital,
  author = {{OECD}},
  title = {Measuring Capital: {OECD} Manual 2009, Second Edition},
  publisher = {OECD Publishing},
  address = {Paris},
  year = {2009},
  doi = {10.1787/9789264068476-en}
}

@book{oecd2018oslo,
  author = {{OECD/Eurostat}},
  title = {Oslo Manual 2018: Guidelines for Collecting, Reporting and Using Data on Innovation},
  edition = {4},
  year = {2018},
  publisher = {OECD Publishing and Eurostat},
  address = {Paris and Luxembourg},
  doi = {10.1787/9789264304604-en}
}

@misc{polyanskiywu2016,
 author = {Polyanskiy, Yury and Wu, Yihong},
 title = {Lecture Notes on Information Theory},
 year = {2016},
 howpublished = {\href{https://ocw.mit.edu/courses/6-441-information-theory-spring-2016/pages/lecture-notes/}{MIT OpenCourseWare}, 6.441, Chapters 2 and 3},
 url = {https://ocw.mit.edu/courses/6-441-information-theory-spring-2016/pages/lecture-notes/}
}

@article{romanelli1994organizational,
  author  = {Romanelli, Elaine and Tushman, Michael L.},
  title   = {Organizational Transformation as Punctuated Equilibrium: An Empirical Test},
  journal = {Academy of Management Journal},
  year    = {1994},
  volume  = {37},
  number  = {5},
  pages   = {1141--1166},
  doi     = {10.2307/256669}
}

@article{rosen1974hedonic,
  author  = {Rosen, Sherwin},
  title   = {Hedonic prices and implicit markets: Product differentiation in pure competition},
  journal = {Journal of Political Economy},
  volume  = {82},
  number  = {1},
  pages   = {34--55},
  year    = {1974}
}

@article{rust1987optimal,
  author  = {Rust, John},
  title   = {Optimal replacement of {GMC} bus engines: An empirical model of {Harold Zurcher}},
  journal = {Econometrica},
  volume  = {55},
  number  = {5},
  pages   = {999--1033},
  year    = {1987}
}

@article{sandborn2008trailing,
  author  = {Sandborn, Peter},
  title   = {Trapped on technology's trailing edge},
  journal = {IEEE Spectrum},
  volume  = {45},
  number  = {4},
  pages   = {42--45, 54, 56--58},
  year    = {2008}
}

@article{saviotti1984theoretical,
  author  = {Saviotti, Pier Paolo and Metcalfe, J. Stanley},
  title   = {A Theoretical Approach to the Construction of Technological Output Indicators},
  journal = {Research Policy},
  year    = {1984},
  volume  = {13},
  number  = {3},
  pages   = {141--151},
  doi     = {10.1016/0048-7333(84)90022-2}
}

@book{saviotti1996variety,
  author    = {Saviotti, Pier Paolo},
  title     = {Technological Evolution, Variety and the Economy},
  year      = {1996},
  publisher = {Edward Elgar},
  address   = {Cheltenham, UK},
  isbn      = {9781852787745}
}

@article{schot2007niches,
  author  = {Schot, Johan and Geels, Frank W.},
  title   = {Niches in Evolutionary Theories of Technical Change: A Critical Survey of the Literature},
  journal = {Journal of Evolutionary Economics},
  year    = {2007},
  volume  = {17},
  pages   = {605--622},
  doi     = {10.1007/s00191-007-0057-5},
  url     = {https://doi.org/10.1007/s00191-007-0057-5}
}

@article{sood2005technological,
  author = {Sood, Ashish and Tellis, Gerard J.},
  title = {Technological Evolution and Radical Innovation},
  journal = {Journal of Marketing},
  year = {2005},
  volume = {69},
  number = {3},
  pages = {152--168},
  doi = {10.1509/jmkg.69.3.152.66361}
}

@article{teece1986profiting,
  author        = {Teece, David J.},
  doi           = {10.1016/0048-7333(86)90027-2},
  journal       = {Research Policy},
  number        = {6},
  pages         = {285--305},
  title         = {Profiting from Technological Innovation: Implications for Integration, Collaboration, Licensing and Public Policy},
  volume        = {15},
  year          = {1986}
}

@article{tripsas2008,
 author = {Tripsas, Mary},
 title = {Customer Preference Discontinuities: A Trigger for Radical Technological Change},
 journal = {Managerial and Decision Economics},
 year = {2008}, volume = {29}, number = {2--3}, pages = {79--97},
 doi = {10.1002/mde.1389}
}

@article{vandenende2005technology,
  author  = {{van den Ende}, Jan and Dolfsma, Wilfred},
  title   = {Technology-Push, Demand-Pull and the Shaping of Technological Paradigms: Patterns in the Development of Computing Technology},
  journal = {Journal of Evolutionary Economics},
  year    = {2005},
  volume  = {15},
  pages   = {83--99},
  doi     = {10.1007/s00191-004-0220-1}
}

@article{windrum2009exploring,
  author  = {Windrum, Paul and Diaz, Cecilia and Filiou, Despoina},
  title   = {Exploring the Relationship between Technical and Service Characteristics},
  journal = {Journal of Evolutionary Economics},
  year    = {2009},
  volume  = {19},
  pages   = {567--588},
  doi     = {10.1007/s00191-009-0143-y}
}

@article{winter1984schumpeterian,
  author  = {Winter, Sidney G.},
  title   = {Schumpeterian Competition in Alternative Technological Regimes},
  journal = {Journal of Economic Behavior \& Organization},
  year    = {1984},
  volume  = {5},
  number  = {3--4},
  pages   = {287--320},
  doi     = {10.1016/0167-2681(84)90004-0}
}

@misc{yinwang2026,
 author = {Yin, Yian and Wang, Dashun},
 title = {Universal Dynamics of Punctuated Progress},
 year = {2026},
 howpublished = {arXiv:2605.16719, version 1},
 url = {https://arxiv.org/abs/2605.16719v1}
}

@article{park2026quantifying,
  title={Quantifying organizational routines: A multidimensional analysis of innovation in the photovoltaic industry},
  author={Park, Hayoung and Jeong, Dawoon and Lee, Jeong-Dong},
  journal={Journal of Evolutionary Economics},
  volume={36},
  number={2},
  pages={34},
  year={2026},
  publisher={Springer}
}

@article{chae2025warning,
  title={Warning signals for long-term innovative potential of genetic research under commercial pressure},
  author={Chae, Junghun and Kim, WooJoong and Jung, Woochul and Jeong, Dawoon and Lim, Roktaek and Chamlagain, Manoj and Jung, Giju and Jang, Juneil and Lee, Jae Won and Kang, Nam Kyu and others},
  journal={bioRxiv},
  pages={2025--02},
  year={2025},
  publisher={Cold Spring Harbor Laboratory}
}

@incollection{kline1986overview,
  author = {Kline, Stephen J. and Rosenberg, Nathan},
  title = {An Overview of Innovation},
  booktitle = {The Positive Sum Strategy: Harnessing Technology for Economic Growth},
  editor = {Landau, Ralph and Rosenberg, Nathan},
  publisher = {National Academy Press},
  address = {Washington, DC},
  year = {1986},
  pages = {275--306},
  doi = {10.17226/612},
  url = {https://www.nationalacademies.org/read/612/chapter/18}
}

@article{bajari2025hedonic,
  author = {Bajari, Patrick and Cen, Zhihao and Chernozhukov, Victor and Manukonda, Manoj and Vijaykumar, Suhas and Wang, Jin and Huerta, Ramon and Li, Junbo and Leng, Ling and Monokroussos, George and Wang, Shan},
  title = {Hedonic Prices and Quality Adjusted Price Indices Powered by {AI}},
  journal = {Journal of Econometrics},
  year = {2025},
  volume = {251},
  pages = {106052},
  doi = {10.1016/j.jeconom.2025.106052},
  url = {https://doi.org/10.1016/j.jeconom.2025.106052}
}
\end{document}